\documentclass[12pt]{article}

 \usepackage{epsf}

\usepackage{amsmath,bm}
\usepackage{amssymb}
\usepackage{graphicx}
\usepackage{amsfonts}         
\usepackage{slashed}

\newcommand{\be}{\begin{equation}}

\newcommand{\ee}{\end{equation}}
\newcommand{\bea}{\begin{eqnarray}}
\newcommand{\eea}{\end{eqnarray}}

\newcommand{\bra}{{\langle}}
\newcommand{\ket}{{\rangle}}

 \newcommand{\myfig}[3]{\begin{figure}[ht]
\begin{center}
\leavevmode \epsfxsize=#2cm \epsfbox{#1}
\end{center}
\caption{#3} \label{fig:#1}
\end{figure}}

\begin{document}

\title{ Scale Invariance, Bounded Rationality and  Non-Equilibrium Economics}

\author{Samuel E. V\'azquez \\
\small{Perimeter Institute for Theoretical Physics, 31 Caroline St. North,}\\
\small{ Waterloo, ON, Canada, N2L-2Y5}}
\date{}

\maketitle

\abstract{We study a class of heterogeneous agent-based models which are based on a basic set of principles, and the most fundamental operations of an economic system: trade and product transformations. 
 A basic guiding principle is scale invariance, which means that  the dynamics of the economy should not depend on the units used to measure the different products. We develop the idea of a ``near-equilibrium" expansion which allow us to study the dynamics of fluctuations around economic equilibrium.  This is similar to the familiar ``perturbation theory" studied in many areas of physics. We study some simple models of both centralized and decentralized markets. We show the relaxation to equilibrium when appropriate. More interestingly, we study a simple model of a decentralized market that shows a spontaneous transition into a monetary phase. We use mean field theory analysis to provide a statistical interpretation of the monetary phase. Furthermore, we show that such phase can be dynamically unstable. Finally, we study some simple centralized financial markets, one of which shows a speculative bubble and a crash.

 %

 }

\section{Introduction}
It is fair to say that the global economic system is one of the most complex structures known in the biosphere.  Compared to physics, economics differs in the important fact that the basic constituents, or ``particles",  are already quite complex: human beings. Moreover, they act on greed and according to expectations about how the future will unfold.

How can one even start to understand such a system when we need to model people's behaviors?  One key revolution came from the concepts of game and decision  theory \cite{gamebook, gamehistory}. In particular, the utility maximization theorem of Morgester and Von Neumann  provided a simple framework in which to model rational decision makers. This lead to the concept of Nash equilibrium, which is basically the core of neoclassical economic theory \cite{macrotheory, samuelson}. 

Nevertheless, the economy is not composed of one or two players, but millions of them. In order to make progress, economist have made many extra assumptions such as the homogeneity of players, perfect information knowledge, unbounded rationality, etc.  In particular, the concept of ``rational expectations" \cite{muth} lead to much progress as it provides closure to the economic equations.  The above concepts usually go under the name of  ``equilibrium theory".  Under these assumptions it is possible to prove beautiful theorems regarding the existence of  Nash equilibria for the whole economy \cite{Arrow}. However, equilibrium theory is not a dynamical theory. Therefore, issues of stability and phase transitions cannot be addressed within this framework. 

 Furthermore, there is an on-going debate on the validity of the assumptions of equilibrium theory \cite{vices}. First, there is the critique to the assumption of the homogeneity of agents. This seems specially concerning in macoeconomic theory, where one usually resorts to the ``representative agent model" \cite{macrotheory}. Quoting Conlisk \cite{conlisk}: ``{\it We model Robinson Crusoe and pretend he's a 7 trillion dollar economy}". In respond to this critique, there has been an increasing number of heterogeneous agent models in the literature (see e.g. \cite{coulon,he,hommes,lebaron}). In fact, it is usually claimed that some of the most salient stylized facts of financial time series are due to  the heterogeneity of agents' strategies. 

Another critique concerns the assumption of unbounded rationality. More precisely, it is usually assumed in most economic theory that agents have perfect knowledge about the economy and the other agents' strategies (this is necessary in order to prove Nash equilibrium). Moreover, they need to maximize their expected utilities with respect to all possible strategies. There is a heated debate about the validity of these assumptions, and their effect on the dynamics of the economy \cite{keiber, rubinstein, aumann, odean}. Models that relax some of these assumptions are usually called under the name of ``bounded rationality".  One of the main critiques to the unbounded rationality paradigm, is the difficult and costly  optimization problems that agents are supposed to do \cite{conlisk}.

In this paper we want to separate the concept of equilibrium from the assumptions of agents' rationality. It is often the case that both concepts get tangled under the same name. Here we say that the economy is in equilibrium if, given the constraints on the system, agents' expected utilities are maximized. In other words, agents are as happy as they get (within their constraints). We will assume the minimal rationality that is required by the Savage-Von-Neumann-Morgenstern axioms of decision theory, which lead to the utility maximization theorem \cite{gamebook}.

The usual approach to go beyond equilibrium theory is to use computational agent based models \cite{agentreview}. Nevertheless, once we break the assumptions of equilibrium theory, one encounters almost an infinite set of possible models. Are there any basic principles in which a semi-realistic economic model should be based on?

  In this paper we introduce a class of economic agent-based models which are built around some basic set of principles and the most fundamental operations of an economic system: trade and transformations.   More precisely, there are a finite number of products in the economy which the agents can trade or transform into other products.  The amount of each product in the economy can be measured in any units the agents want to use. One of the basic principles  is that the dynamics of the economy should be invariant under a change of product measure. This principle is analogous to the ``gauge symmetries" found in physical systems\footnote{The models that we will study turn out to have an extra ``accidental" gauge symmetry which we discuss below.}. We call this gauge symmetry {\it scale invariance}\footnote{The idea that gauge theories might have applications in economics was introduced to me by Lee Smolin \cite{Smolin}. Nevertheless, this idea has been considered before by Ilinsky \cite{Ilinsky},  Malaney and Weinstein  \cite{Malaney}, which have also been influential in Smolin's ideas.}.  We will show that such a symmetry principle  gives rise to very useful restrictions on the possible dynamics.  
  
  Another important guiding principle is {\it bounded rationality}. Our agents are boundedly rational in that they do not need to know all the information of the economy in order to make a decision. Moreover, they only have a finite set of possible operations that they can do at every time step. They choose to do the operations which give them maximum satisfaction.

  Another important principle is that prices are relational and they simply describe the relative flux of products between two or more agents. Such fluxes do not have to be the same among all agents.     Therefore, not all agents have to agree on a ``price". Related to this fact, we assume that 
  agents possess a  ``what-by-what" matrix which represent the agents' opinion on the exchange rate between the different products in the economy.  These represent part of the information that agents have about the economy. 
  This type of economic modeling is based in the ideas of Brown et. al \cite{partecon}.  The framework presented here allows to simulate heterogeneous agents expectations. 

We develop the idea of a ``near-equilibrium" state, where the agents' expected utility is close to being maximal. In this way we are able to derive approximate ``flow" equations which describe how many products are exchanged or transformed in any of the possible economic operations. The equations are written for general expected utilities, and they allow for fast numerical implementation.  Moreover, our semi-analytic approach allow us to apply mean-field theory to certain agent-based models. We argue that the concept of a ``near-equilibrium" state is similar to the familiar ``perturbation theory" of Quantum Field Theory, in which we expand the theory around a particular vacuum. This provides a constructive way to study fluctuations around an economic equilibrium. 

We then present various numerical simulations of particular market models. We study both decentralized and centralized markets.  One of the interesting questions we address is the emergence of (commodity) money.  We study a class of decentralized markets that show a spontaneous transition into a monetary economy. In other words, agents collectively trade using a common product as a ``currency". We use a mean field theory analysis to show that such transition has a purely statistical origin. Furthermore, we show that in some cases such monetary phase can be dynamically unstable.

We also give simple examples of centralized financial markets. In one of these models, we include agents' speculations about the future price of the products, which leads to a bubble and a crash.

The organization of the paper is as follows. In section 2, we introduce a list of basic principles that we use as a guidance to build our class of agent-based models. In section 3, we elaborate on the allowed form of the utility functions. In section 4 we introduce the idea of a near-equilibrium state, and derive a set of propositions which give us the flow equations used in the numerical algorithms. We also make an analogy with fluid dynamics.  In section 5 we present numerical simulations of decentralized markets. The monetary phase is studied in section 5.4. In section 6 we present simulations of centralized markets. In section 7 we conclude.

\section{A Minimal Economic System }
In this section we present the basic properties and assumptions of the ``minimal" economic system we wish to study. Some of the statement that we will make are very generic and need to be refined in order to construct particular models.  This will be done in the next sections.

\begin{itemize}
\item {\it Agents and Products}: We will consider a fixed number of agents $N$, which we label by Greek letters $\alpha ,\beta = 1,2,\ldots, N$.  The set of all such agents is denoted by ${\cal A}$, and we will use the notation $\alpha \in {\cal A}$ to refer to some generic agent $\alpha$ in the set ${\cal A}$. There will be $P$ different kinds of products which we label with Latin letters, $i,j = 1,2, \ldots, P$. The set of all kinds of products is denoted by ${\cal P}$. By abuse of notation, we will sometimes refer to an arbitrary product by the latin letter itself, e.g. $i \in {\cal P}$.

\item {\it Time}: We assume that the economy evolves in integer time steps. 
\item  {\it Inventories and Scale Invariance}: Every agent has an inventory of products, with quantity $n^i_\alpha$ of product $i$. Sometimes we will write $\vec n_\alpha = (n^1_\alpha, n^2_\alpha, \ldots )$. 
 Agents are free to measure each product in arbitrary units.  For example, product 1 might be measured in liters, and product 2 in pounds. In order to make a trade (see below) the two agents must agree on a unit of measure. Therefore, for simplicity we assume that all agents agree on the same units of measure for each product. Nevertheless, we assume that the dynamics of the economy is invariant under changes of units for all agents. Under such change, the inventories transform as,
 \be \label{gauge} n^i_\alpha \rightarrow \phi^i n^i_\alpha\;,\;\;\; \phi^i \in \mathbb{R}_+\;,\;\; \forall \alpha\; \in {\cal A}\;.\ee
We say that the transformation (\ref{gauge}) is a ``gauge" symmetry. We call this symmetry {\it scale invariance}\footnote{In this paper we only study the case of time-independent scale invariance. That is, $\phi^i$ in Eq. (\ref{gauge}) is a constant in time. The more general case of a time dependent scale invariance will be studied elsewhere.}.
  
\item {\it State Space}:  The state of the economy  at any point in time is given by the set  $\psi= \{n^i_\alpha |  i \in {\cal P},\;  \alpha \in {\cal A}\}$. Let ${\cal H}$ the set of all such possible states. We see that for fixed agent and product number, and if we do not allow agents to have negative amount of products, 
\be {\cal H} \simeq (\mathbb{R}_+)^{P  \times N}\;.\ee

\item {\it Agents' Rationality}: We assume that agents have beliefs about how the future will unfold. Different agent can have different beliefs and future time frames in mind. We will denote the belief of agent $\alpha$ by ${\cal I}_\alpha$.  By abuse of notation we will sometimes call such belief  ``information". Nevertheless, we should stress that such information might be biased or might not be based on a good model of Nature. Furthermore, agents are not assumed to have full knowledge of the state of the economy and other agents' beliefs. Therefore, we do not assume {\it rational expectations}. 

We assume, however,  that agents make {\it rational decisions} given the information ${\cal I}_\alpha$. In other words, we assume that agents' decisions satisfy the Savage-Von-Neumann-Morgenstern axioms of decision theory \cite{gamebook}. Then, by the Utility Maximization Theorem, it follows that such agents must carry an index of satisfaction $\Omega_\alpha$, $\alpha \in {\cal A}$. The index of satisfaction is a map,
\be \Omega_\alpha : \vec{n}_\alpha \times {\cal I}_\alpha \rightarrow \mathbb{R}\;.\ee
Agents always seek to maximize $\Omega_\alpha$. 
We assume the following properties of the index of satisfaction,
\be\label{indexconditions} \partial_i \Omega_\alpha > 0 \;,\;\;\; \partial_i^2 \Omega_\alpha  < 0 \;,\;\; \forall \; i \in {\cal P}  \ee
where $\partial_i = \partial/\partial n^i_\alpha$. 

The index of satisfaction is given in terms of the agents' utility $U_\alpha$ as,
\be \Omega_\alpha   = \text{E}\left[ U_\alpha | {\cal I}_\alpha\right]\;.\ee
where $\text{E}[ \cdot| {\cal I}_\alpha]$ denotes the future expectation given the information ${\cal I}_\alpha$. The utility also satisfies the properties (\ref{indexconditions}).

\item {\it What-by-What Matrices}: At any point in time, some agents might have an opinion on what each product of their inventory is worth in units of another product\footnote{As we will discuss in the next section, not all agents necesarily have such an opinion. Moreover, some agents might have a w-w matrix with entries for only a particular subset of the products that he/she possesses.}. That is, they carry an exchange rate matrix which we call ``what-by-what matrix" (w-w matrix for short). The entries of such matrix are denoted by $(M_\alpha)^i_{\;j}$.    Note that this matrix is part of the {\it beliefs} of the agent, and so by abuse of notation we could say that $M_\alpha \in {\cal I}_\alpha$. Moreover, agents not only have an idea of what $M_\alpha$ should be today but what it should be in the future based, perhaps, in some previous experience. 

Let $\cal P_\alpha$ be the set of products for which agent $\alpha$ has an entry in his/her w-w matrix.
Rationality of the agent requires that the w-w matrix obeys the reciprocal relation,
\be \label{cons1}(M_\alpha)^i_{\;j}  = \frac{1}{(M_\alpha)^j_{\;i}}\;,\;\;\; \forall\; i,j \in {\cal P_\alpha}\;,\ee
and the transitivity conditions,
\be\label{cons2} (M_\alpha)^i_{\;j}  = (M_\alpha)^i_{\;k} (M_\alpha)^k_{\;j}\;,\;\;\; \forall \; i,j,k \in {\cal P_\alpha}\;.\ee
We say that  a w-w matrix that obeys the conditions (\ref{cons1}) and (\ref{cons2}) is  {\it consistent}.
Under the gauge transformation (\ref{gauge}), the w-w matrix transforms as,
\be (M_\alpha)^i_{\;j}  \rightarrow  \phi^i (M_\alpha)^i_{\;j}  (\phi^j)^{-1}\;.\ee

\item {\it Scale Invariance and Utility}: Given the structures given so far, we note that all tensors must be constructed out of the inventories $n_\alpha^i$ and the w-w matrices. Moreover, scale invariance dictates their transformation properties with respect to (\ref{gauge}). For example, a tensor of the form $T^{ij}_{\;k}$ must transform as $T^{ij}_{\;k} \rightarrow \phi^i \phi^j (\phi^k)^{-1} T^{ij}_{\;k}$. 
Therefore, one cannot have a combination such as  $n^1_\alpha + n^2_\alpha$, since $n^1_\alpha$ and $n^2_\alpha$ are measured in different units. One can have, however,  $T^1 = n^1_\alpha  + (M_\alpha)^1_{\;2} n^2_\alpha$. It is easy to see that this will transform as $T^1 \rightarrow \phi^1 T^1$ under (\ref{gauge}).

The form of the utility function is also restricted by scale invariance. First, let's suppose that $U_\alpha$ has a fixed scale dependence with respect to the transformations (\ref{gauge}). Then, consider the difference in happiness between an allocation $\vec n_\alpha '$ and $\vec n_\alpha$:
\be \Delta \Omega_\alpha  = \Omega_\alpha(\vec n_\alpha ') - \Omega_\alpha(\vec n_\alpha)\;.\ee
This is certainly a physically meaningful quantity which allows the agent to determine if $\vec n_\alpha '$ is more desirable than $\vec n_\alpha$. 
Doing the transformation (\ref{gauge}), it follows that the difference in happiness {\it for the same physical allocations} $\vec n_\alpha '$  and $\vec n_\alpha$, will depend on the units used to measure the products.  This does not make any sense\footnote{It is like saying:  ``I would get more happiness if I buy the same amount of  gas at the same price but using Liters instead of Gallons".}. Therefore, to avoid such paradoxes, the utility function much change at most by a constant under (\ref{gauge}). It follows that,
\be U_\alpha = \log \tilde  U_\alpha\;,\ee
where $\tilde  U_\alpha$ has a fixed scale dependence under (\ref{gauge}). Scale invariance and the consistency of the w-w matrix,  will imply a bigger gauge symmetry which we will discuss in the next section.

\item {\it Trading Network}: We fix the ``background" market structure which includes the trading network and the operations which each agent are allow to make (see below). We will consider the case of both a decentralized and centralized market.

\item {\it Economic Operations}: An operation $\hat {\cal O}$ on the market is a map which changes agents' inventories. Therefore, it is also a map on the economic state as:
\be \hat {\cal O} : {\cal H} \rightarrow {\cal H}\;.\ee
We consider two basic kinds of operations: a {\it trade} ($ \hat {\cal T}$) and a {\it transformation} or {\it metabolism} ($\hat {\cal M}$). A trade is an exchange of products between two agents. Therefore, product number is conserved in this case. A metabolism is a transformation of the form,
\be i \rightarrow j\;.\ee
Such transformation can be synchronized into more complex ones, e.g.
\be (i\rightarrow k) + (j \rightarrow k) = i + j \rightarrow k \;.\ee
We also allow the possibility of a {\it decay} $({\cal D})$ such that 
\be i \rightarrow E\;,\ee
where $E$ is the energy dissipated in the process. The dissipated energy cannot be used by the agents and so it cannot lead to a change in utlity. Nevertheless, the fact that decays reduce product number, implies that utility will be reduced by a decay process.

 Since economic operations affect the agents' inventory, they also cause a change in their satisfaction. Let,
\be \Delta_{\cal O} \Omega_\alpha = \left.\left[(\hat {\cal O} \Omega_\alpha) - \Omega_\alpha\right] \right|_{{\cal I}_\alpha}\;,\ee
for a general operation $\hat {\cal O}$. 
In other words, $\Delta_{\cal O} \Omega_\alpha$ is the change of the agent satisfaction due to the operation $\hat {\cal O}$ with fixed information.

\item {\it Dynamics}: Given the market structure restrictions,  at every time step each agent will perform the operation $\hat{\cal O}$ that give him/her the maximum positive change in satisfaction $\Delta_{\cal O} \Omega_\alpha > 0$.  
As we pointed out, decay processes will have $\Delta_{\cal D} \Omega_\alpha < 0$. Nevertheless, we assume that such processes are beyond the control of the agents. 

We wish to emphasize that agents will not do inter-temporal optimization problems such as the ones considered in discount factor models \cite{macrotheory}.  Instead, they make a decision at every time step regarding which operation maximizes his/her satisfaction.

\item {\it Equilibrium}: We define equilibrium as a state where there is no operation $\hat {\cal O}$ such that $\Delta_{\cal O} \Omega_\alpha > 0$. In other words, all agents are as satisfied as they get {\it given the operations they can perform}.  We will be more explicit about the meaning of equilibrium for each one of the economic operations, in  section 4. 

\item {\it Linear Response Theory and the Near-Equilibrium Expansion}: We will derive approximate expressions for the action of each of the transformations given that agents are close to being satisfied. Thus, $\Delta_{\cal O} \Omega_\alpha \gtrsim 0$

\end{itemize}

The properties given above define a broad class of minimal economic systems.  
However, in order to produce particular models we need to be more specific with respect to the market structure and the way agents maximize their satisfactions. In this minimal model we are considering anonymous agents, and so we do not have contracts and credit. The addition of such features will be left for future work. Nevertheless, one can introduce interest rates as an exogenous process if desired.

One of the biggest critiques of the  unbounded rationality paradigm is the complexity of the optimization processes that agents must do in order to find the equilibrium \cite{conlisk}. This also presents a challenge when it comes to numerical simulations: utility maximization is a very inefficient algorithm specially if we want to have heterogeneous agents expectations. In this paper we solve this problem in two steps. First, we go back to the very basic operations of economics which are {\it trade} and {\it metabolisms}. In this way, at every time step agents have only a finite number of operations that they can do. This is, in fact, more akin to real life.  Second, and more importantly, we will develop a ``linear response theory" such that satisfaction maximization can be approximated in closed form for general utilities, assuming that agents are close to being satisfied. That is, agents make decisions following ``gradients" of satisfaction.

This kind of approximation is reminiscent of the familiar ``perturbation theory" used in particle and other areas of physics. In perturbation theory, one starts with a particular vacuum state of the theory. In economic terms, this means first identifying a particular equilibrium state (or a family of them). One can then do a systematic expansion of small perturbations around such equilibrium. The study of such perturbations is necessary to gain an understanding of the stability of the vacuum state (or economic equilibrium).

Another analogy with physics comes from fluid dynamics where one assumes a near-equilibrium state at every point in the fluid. Then, one makes an expansion in gradients of velocity, temperature and chemical potential that describe the small deviations from local equilibrium. From this point of view, and taking into account the principles outlined above we can say that:
 {\noindent \it The physics of economics is the study of the flow of products through the social network in respond to ``gradients" of satisfaction. Moreover, prices are relational and given by the ratio of fluxes: e.g. $M^i_{\;j} = - \Delta n^i/\Delta n^j$ for a trade $i\leftrightarrow j$}.

In the next section, we will examine in more detail how scale invariance restricts the form of the utility functions. 

\section{Scale Invariant Utility Functions}
In physics, gauge symmetries can greatly reduce the number of parameters in a physical model. In our scale-invariant formulation of economics, this is also the case. 
As we noted in the previous section, scale invariance severely restricts the form of the utility functions that agents can use.  In this section we will study some examples of scale invariant utility functions and their corresponding index of satisfaction. Nevertheless, we stress that the ``flow equations" derived in the next section are written for a general index of satisfaction.

An example of a scale invariant utility which we use below is,
\be \label{util1} U_\alpha = \log\left[ \sum_{i \in {\cal P}} (n^i_\alpha)^\nu [(M_\alpha)^j_{\;i}]^\nu\right]\;,\ee
where $0 < \nu < 1$, and the index $j$ is arbitrary.  The exponent $\nu$ measures how much weight the agent puts on the value of his w-w matrix vs. his intrinsic  inventory preference. For example, for $\nu = 1$, the utility is simply the log of the inventory, and so the agent is a wealth maximizer. On the other hand, for small $\nu$ we will see that the agents put more attention to their number of products $\vec{n}_\alpha$.  Also note that $\partial_i \partial_j U_\alpha \neq 0$ and so all products are substitutes for this agent.

The argument of the log is being measured in units of product $j$. Now suppose we made some other choice instead of $j$. Using the consistency of the w-w matrix one finds,
\be \log\left[ \sum_{i \in {\cal P}} (n^i_\alpha)^\nu [(M_\alpha)^j_{\;i}]^\nu\right] = \log\left( [(M_\alpha)^j_{\;k}]^\nu\right) + \log\left[ \sum_{i \in {\cal P}} (n^i_\alpha)^\nu [(M_\alpha)^k_{\;i}]^\nu\right]\;.\ee
Therefore, the utility (and hence the index of satisfaction) changes by a constant. Since all economic operations involve derivatives of the utility, we see that the choice of measure is arbitrary. 
This is an accidental gauge symmetry.  With this utility, every product is treated in equal footing, and so that the emergence of money will be a purely dynamical phenomenon. In fact, the monetary phase can be interpreted as a breaking of this gauge symmetry.

An important property of the utility (\ref{util1}) is that, for $\nu \neq 0,1$,  
\be\label{convex} \partial_i U_\alpha \rightarrow \infty\;,\;\; \text{as}\;\;\; n^i_\alpha \rightarrow 0\;.\ee
This property, along with Eqs. (\ref{indexconditions}) implies that the utility function considered above (and hence the index of satisfaction)  is {\it convex}. 
Convexity promotes diversification\footnote{The fact that convexity implies diversification is an important fact used in modern portfolio theory \cite{Atilio}.}. 
 For example, an agent will be happier having two kinds of products that only one. To see this, consider an agent with product $i$ and $j$. Furthermore, suppose that it has the same net value in each product: $n^i_\alpha = (M_\alpha)^i_{\;j} n^j_\alpha$. Then,
\be \log \left[ (n^i_\alpha)^\nu + \left[ (M_\alpha)^i_{\;j} n^j_\alpha \right]^\nu\right] = \log \left[ 2 (n^i_\alpha)^\nu \right] > \log \left[ (2 n^i_\alpha)^\nu \right]\;.\ee

Convexity might seem like a special property of this utility function. However, it serves a deeper purpose. A purely wealth maximizer agent will tend to give away all his ``money" in a single transaction. This is obviously not a realistic human behavior. In macroeconomic theory it is often said that the fundamental problem that agents face is, precisely, how much to consume and how much to save to consume later \cite{macrotheory}. It is then assumed that agents do inter-temporal optimization, and thus somehow they must know their consumption plan for the rest of their lives! This is a very unrealistic assumption. 

In our case, one can avoid facing this problem by considering convex utilities. In this way, the agent will not be willing to give up all of any of the products he/she possesses. Convexity will also be important in order to prove some of the propositions in the next section. Therefore, for the most part of this paper we will only consider convex utility functions. 
Note that we can always take $\nu  \lesssim 1$ in order to have approximate wealth maximizer. The main results are insensitive to wether $\nu$ is exactly one or not. 

Utility functions can also be completely independent on the w-w matrices, e.g.
\be U_\alpha = \sum_{i \in {\cal P}} \log n^i_\alpha\;.\ee
In this case $\partial_i \partial_j U_\alpha = 0$, and so the products are not substitutes.  Utility functions can also depend on both substitutes and non-substitutes.

\section{Linear Response and Economic Operations}
In this section we will explain in more detail the different kinds of economic operations that agents are allowed to do. The operations that we will consider are bartering between two agents, an agent making a trade at fixed exchange rate, a production process and a central agent that sets prices in respond to excess demand. In sections 5 and 6, we will explain how to use these operations in particular market models.  Our main goal here is to derive approximate formulas for the flow of products trough the economy given that agents are trying to maximize their satisfaction. These formulae will be implemented below both in numerical simulations, and in mean-field theory analysis. The prevailing theme will be the near-equilibrium expansion.

\subsection{Trading at a fixed Exchange Rate}
In this section we consider an agent that exchanges two products $i,j$  at a fixed rate $M^i_{\;j}$ which might be different from his own w-w matrix.  More specifically, the trading process which we call $\hat{\cal T}_1$ takes the form,
\be \hat{\cal T}_1: \;\;\;\;n^i_\alpha \rightarrow n^i_\alpha + \Delta n^i\;,\;\;\; n^j_\alpha \rightarrow n^j_\alpha - M^j_{\;i} \Delta n^i\;.\ee
Our goal is to find $\Delta n^i$, given that the agent is trying to maximize his satisfaction. We assume that the agent is in a near-equilibrium state, which we define in detail below. The proof of the following proposition will require a convex index of satisfaction.

\bigbreak
\noindent{\bf Proposition 1:} {\it In a near-equilibrium state, the amount of product $i$ and $j$ that agent $\alpha$ would trade at a fixed rate $M^j_{\;i}$ is given by} 
\be  \label{dni} \hat {\cal T}_1:\;\;\; \Delta n^i \approx J^{ii}_\alpha \left(\partial_i \Omega_\alpha - M^j_{\;i} \partial_j \Omega_\alpha\right)\;,\;\;\; \Delta n^j = - M^j_{\;i }\Delta n^i \;,\ee
{\it where}
\be \label{J}  J^{ii}_\alpha  = (\partial_j\Omega_\alpha)^2 \left[ 2 \partial_i \Omega_\alpha \partial_j\Omega_\alpha \partial_i\partial_j\Omega_\alpha - \partial_i^2 \Omega_\alpha (\partial_j\Omega_\alpha)^2 - \partial_j^2 \Omega_\alpha (\partial_i\Omega_\alpha)^2\right]^{-1}\;. \ee
{\it Moreover, the gain in satisfaction is given by,}
\be \label{dOmegaT1} \Delta_{{\cal T}_1} \Omega_\alpha \approx \frac{1}{2} J^{ii}_\alpha \left(\partial_i \Omega_\alpha - M^j_{\;i} \partial_j \Omega_\alpha\right)^2\;.\ee
{\it The trade $\hat{\cal T}_1$ is allowed iff $J^{ii}_\alpha > 0$}.

\bigbreak
\noindent {\bf Proof : } In order to prove this proposition we need to ask ourselves: what does it mean to be in equilibrium? The maximum increase in happiness is a change of inventory $\Delta n^i_*$ such that,
\be \label{max1} \left. \frac{\partial}{\partial \Delta n^i} (\Delta_{{\cal T}_1} \Omega_\alpha) \right|_{\Delta n^i_*}= \left. \left(\partial_i \Omega_\alpha - M^j_{\;i} \partial_j \Omega_\alpha\right) \right|_{\Delta n^i_*} = 0\;, \;\;\; \left.\frac{\partial^2}{(\partial \Delta n^i)^2} (\Delta_{{\cal T}_1} \Omega_\alpha)\right|_{\Delta n^i_*} < 0\;,\ee
where the satisfaction index is now a function of $\Delta n^i$ as,
\be \Omega_\alpha = \Omega_\alpha(n^i_\alpha + \Delta n^i, n^j_\alpha - M^j_{\; i} \Delta n^i)\;.\ee
We say that the agent is at equilibrium if the solution to Eq.(\ref{max1}) is $\Delta n^i_* = 0$. 

That a point satisfying the first equation in (\ref{max1}) exist, follows from the convexity of the index of satisfaction. To see this, suppose that $\partial\Omega_\alpha/\partial(\Delta n^i) = (\partial_i \Omega_\alpha - M^j_{\;i} \partial_j \Omega_\alpha) < 0$. Then, we can make $\Delta n^i < 0$, and keep depleting the amount of product $i$. However, the convexity of the index of satisfaction implies that  $\partial_i \Omega_\alpha$ will eventually become very large and positive. Thus, the sign of  $\partial\Omega_\alpha/\partial(\Delta n^i)$ will be reversed. By continuity, there must be a value of $\Delta n^i$ where $(\partial_i \Omega_\alpha - M^j_{\;i} \partial_j \Omega_\alpha) = 0$. Whether this is a minimum or maximum must be checked by computing the second derivative w.r.t. $\Delta n^i$. 

Therefore, near equilibrium we can expand the gain in satisfaction as,
\bea \label{dOT1} \Delta_{{\cal T}_1}  \Omega_\alpha &\approx& \Delta n^i  \left. \left(\partial_i \Omega_\alpha - M^j_{\;i} \partial_j \Omega_\alpha\right)\right|_{\Delta n^i = 0} \nonumber \\
&&+  \frac{1}{2} (\Delta n^i)^2   \left. \left[ \partial_i^2 \Omega_\alpha + \left (M^j_{\;i}\right)^2 \partial_j^2\Omega_\alpha - 2 M^j_{\;i} \partial_i \partial_j \Omega_\alpha\right]\right|_{\Delta n^i = 0}  + \ldots \eea
The maximization equation reads
\be \frac{\partial}{\partial \Delta n^i} (\Delta_{{\cal T}_1} \Omega_\alpha) = 0\;.\ee
The linearized solution is given by Eq. (\ref{dni}), where we evaluate all quantities at  $\Delta n^i = 0$. Note that all indices in Eq. (\ref{dni}) reflect their correct tensor transformation properties under Eq. (\ref{gauge}). In writing Eq. (\ref{J}) we have used the fact that at equilibrium, the inventory of agent $\alpha$ will obey $M^j_{\;i} = \partial_i\Omega_\alpha/\partial_j\Omega_\alpha |_\text{eq}$. Therefore, we can approximate, $M^j_{\;j} \approx \partial_i\Omega_\alpha/\partial_j\Omega_\alpha  + {\cal O}(\delta)$, where $\delta$ means a first order deviation from equilibrium.
The gain in satisfaction can be computed inserting (\ref{dni}) in (\ref{dOT1}), and it is given by Eq. (\ref{dOmegaT1}). 

We see that Eq. (\ref{dni}) represent a maximum {\it iff} $J^{ii}_\alpha > 0$. If $J^{ii}_\alpha <0$ the trade will not proceed as it will represent a decrease in satisfaction.  For most utilities, since $\partial_i^2 \Omega_\alpha < 0$, it is easy to satisfy $J^{ii}_\alpha > 0$. Nevertheless, we need to check case by case since one can have $\partial_i \partial_j \Omega_\alpha < 0$. 

\noindent$\square$

In sections 5 and 6 we will discuss how the agents use the result of Proposition 1 to make a decision of wether to trade with $\hat {\cal T}_1$ or not. In any case, once the agent decides that he will trade with the operation $\hat {\cal T}_1$, Eq. (\ref{dni}) give us the flow of objects within the linear response approximation. This should be compared to the constitutive relations of hydrodynamics where the shear tensor has an expansion in gradients of the fluid velocity with the first term being the well-known ``shear viscosity". In our case, $J^{ii}_\alpha$ plays a similar role as a ``transport coefficient" that tell us how the products will flow under a gradient of ``satisfaction". Note that this proposition can be easily extended to higher orders.

\subsection{Barter}
In a barter economy, two agents  must bargain an exchange rate and how much product to trade. We will call such operation $\hat {\cal T}_2$. It is easy to get carried on trying to come up with a complicated algorithm to describe the bargaining process. However, one can simply assume that both agents ``quickly" converge to an equilibrium. The reader might ask: isn't this going back to general equilibrium theory, which is what we are trying to avoid? To answer this question, one must consider what are the different time scales involved in the problem. How much time does it take for two parties to reach a price agreement vs. the time it takes the whole economy to equilibrate (if it does)? We can safely assume that two agents can reach an agreement much faster than the relaxation time of the whole economy. 

This situation should, again, be compared to the case of fluid dynamics. The fact that we can speak of a temperature of a fluid at some point in space, means that we are assuming the fluid is in a state of (near) equilibrium {\it locally} in space. However, the global dynamics of the fluid can be quite complicated (e.g. turbulence).

After the bargaining process, the inventories of the two agents $\alpha$ and $\beta$ will be updated as,
\be n^i_\alpha \rightarrow n^i_\alpha + \Delta n^i\;,\;\;\; n^j_\alpha \rightarrow n^j_\alpha + \Delta n^j\;,\;\;\; n^i_\beta \rightarrow n^i_\beta - \Delta n^i\;,\;\;\; n^i_\alpha \rightarrow n^j_\beta - \Delta n^j\;.\ee
When both agents are in a near equilibrium state, there is a unique two-agent equilibrium which is summarized in the following proposition:
\bigbreak

\noindent {\bf Proposition 2 :} {\it Near equilibrium, the amount of product $i$ and $j$ that two agents $\alpha$ and $\beta$ exchange as a result of a bargaining process ($\hat {\cal T}_2$) is given by,}
\bea \label{dniT2}\Delta n^i &\approx& L^{i i j} \left(\partial_i \delta \Omega \partial_j \Omega - \partial_i \Omega\partial_j \delta\Omega\right) + {\cal O}(\delta \Omega^3)\;, \\
\label{dnjT2}\Delta n^j &\approx& -\frac{\partial_i \Omega}{\partial_j \Omega} \Delta n^i + {\cal O}(\delta \Omega^2)\;.\eea
{\it where}
\be\label{Liij} {L^{iij}  = - (\partial_j \Omega) \left[ \partial_i^2 \Omega (\partial_j\Omega)^2+\partial_j^2 \Omega (\partial_i\Omega)^2 - 2 \partial_i \Omega \partial_j \Omega \partial_i \partial_j \Omega  + {\cal O}(\delta \Omega^2) \right]^{-1} + {\cal O}(\delta\Omega^2)\;.}\ee
{\it and}
\be \Omega \equiv \Omega_\alpha + \Omega_\beta\;,\;\;\; \delta \Omega \equiv \Omega_\alpha - \Omega_\beta\;.\ee
{\it Moreover, the gain in satisfaction is given by,}
\be{ \label{dOmegaT2} \Delta_{{\cal T}_2}\Omega_\alpha \approx\Delta_{{\cal T}_2}\Omega_\beta \approx \frac{ L^{iij}}{4\partial_j \Omega}  \left(\partial_i \delta \Omega \partial_j \Omega - \partial_i \Omega\partial_j \delta\Omega\right) ^2\;.}\ee
{\it The trade $\hat{\cal T}_2$ will proceed iff $L^{iij} > 0$}.

\bigbreak
\noindent {\bf Proof:} 
To prove this proposition we can simply use the fact that at equilibrium, by definition, the excess demand will be zero. If agent $\alpha$ and $\beta$ are in a near-equilibrium state, we have already calculated the amount of goods that they require to maximize their satisfaction (proposition 1). In terms of $i$ we have,
\be\label{dnialpha} \Delta n^i_\alpha \approx J^{ii}_\alpha \left(\partial_i \Omega_\alpha - (M_*)^j_{\;i} \partial_j \Omega_\alpha\right)\;,\;\;\; \Delta n^i_\beta \approx J^{ii}_\beta \left(\partial_i \Omega_\beta - (M_*)^j_{\;i} \partial_j \Omega_\beta\right)\;,\ee
where $M_*$ is the equilibrium exchange rate, which is to be determined.

The condition of zero excess demand is simply $\Delta n^i_\alpha + \Delta n^i_\beta = 0$. This determines the exchange rate to be,
\be \label{Mstar} (M_*)^j_{\;i} = \frac{J^{ii}_\alpha \partial_i \Omega_\alpha + J^{ii}_\beta \partial_i \Omega_\beta}{J^{ii}_\alpha \partial_j \Omega_\alpha + J^{ii}_\beta \partial_j \Omega_\beta}\;.\ee
Inserting Eq. (\ref{Mstar}) this back into (\ref{dnialpha}) gives,
\be \label{dnialpha1} \Delta n^i_\alpha  \approx \frac{J_\alpha^{ii} J_\beta^{ii}}{ J_\alpha^{ii}  \partial_j \Omega_\alpha+J_\beta^{ii}  \partial_j \Omega_\beta} \left[\partial_i\Omega_\alpha \partial_j\Omega_\beta - \partial_i \Omega_\beta \partial_j\Omega_\alpha\right]\;. \ee

Since we have been using the near-equilibrium approximation in (\ref{dnialpha}), we must be consistent with it. Looking at Eq. (\ref{dnialpha1}), we notice that if both agents have the same preferences, there will be no trade. More precisely, $\Delta n^i_\alpha = 0$ if $\partial_i(\Omega_\alpha - \Omega_\beta) = 0$ and $\partial_j(\Omega_\alpha - \Omega_\beta) = 0$. This makes perfect sense: the only reason two agents will ever trade is because they have different preferences (or expectations). To be consistent with the near-equilibrium approximation, we should assume that the two agents' preferences are not very different. To do this, we define,
\be \Omega = \Omega_\alpha + \Omega_\beta\;,\;\;\; \delta \Omega = \Omega_\alpha - \Omega_\beta\;.\ee
We now need to expand to leading non-trivial order in (derivatives of) $\delta \Omega$. 

Thus, to leading order, the exchange rate (\ref{Mstar}) gives,
\be \label{Mstar1} (M_*)^j_{\;i}  \approx \frac{\partial_i \Omega}{\partial_j\Omega} + {\cal O}(\delta \Omega^2)\;.\ee
Moreover, Eq. (\ref{dnialpha1}) gives precisely Eq. (\ref{dniT2}) of the proposition.

The gain in satisfaction for both agents can be calculated from Eq. (\ref{dOmegaT1}) of proposition 1. Inserting Eq. (\ref{Mstar1}) into (\ref{dOmegaT1}) and respecting the near-equilibrium approximation, we get Eq. (\ref{dOmegaT2}).

\noindent$\square$

It is interesting that to the leading approximation, the two-agent equilibrium is egalitarian ($\Delta \Omega_\alpha = \Delta \Omega_\beta$). One can then, alternatively, prove this proposition starting from the assumption that the two-agent equilibrium is egalitarian. For an egalitarian equilibrium there is a unique way to decide what to trade:  the agents simply trade the pair which gives them the best (mutual) satisfaction. If one goes beyond the near-equilibrium approximations used above, both agents will not necessarily get the same satisfaction.

\subsection{Metabolisms}
In this section we will consider a transformation or metabolism of the form,
\be \hat {\cal M}:\;\;\; i + j \rightarrow k\;.\ee
Such transformation will be made using the current technology which  we encode in a {\it production matrix}.

\bigbreak
\noindent {\bf Definition 1:} {\it A  Production Matrix is the rate into which product $i$ and product $j$ are depleted in the conversion $i + j \rightarrow k$. We write}
\be \frac{dn^i}{dn^k} = - P^i_{\;k}\;,\;\; \frac{dn^j}{dn^k} = - P^j_{\;k}\;,\;\;\; P^i_{\;k}, P^j_{\;k} > 0\;,\ee
{\it and we assume that both products $i$ and $j$ must be present at the same time. By convention, we will always take the upper index to be one of the input products, and the lower index the output product. Thus, note that $P^i_{\;k} \neq 1/P^k_{\;i}$ in general, as $P^i_{\;k}$ and $P^k_{\;i}$ represent very different processes.  }

Agents will only perform metabolisms that increase their satisfaction. 
We are now ready to calculate how much output is produced in a production process, assuming  near-equilibrium state.

\bigbreak
\noindent {\bf Proposition 3 :} {\it Near equilibrium and with constant production matrices, the output of a metabolic process $\hat {\cal M} : \; i + j \rightarrow k$ can be approximated by }
\bea \label{dnk}\Delta n^k &\approx&   K^{kk}_\alpha \left( -\partial_i \Omega_\alpha P^i_{\;k} -\partial_j \Omega_\alpha P^j_{\;k}  + \partial_k\Omega_\alpha\right)\;, \\
\label{dniM}\Delta n^i & = &- P^i_{\;k} \Delta n^k\;,\\
\label{dnjM}\Delta n^j &=&  -P^j_{\;k} \Delta n^k\;,\eea
{\it where}
\bea K^{kk}_\alpha  &=&  -\left[ \partial_i^2 \Omega_\alpha (P^i_{\;k})^2 +\partial_j^2 \Omega_\alpha (P^j_{\;k})^2   + \partial_k^2 \Omega_\alpha + 2 \partial_i \partial_j \Omega_\alpha P^i_{\;k} P^j_{\;k} \right.\nonumber \\
&& \left.  - 2 \partial_i \partial_k \Omega_\alpha P^i_{\;k} - 2 \partial_j \partial_k \Omega_\alpha  P^j_{\;k}\right]^{-1}\;.\eea
{\it The change in satisfaction is given by,}
\be { \label{dOM} \Delta_{\cal M} \Omega_\alpha \approx  \frac{1}{2}  K^{kk}_\alpha \left( -\partial_i \Omega_\alpha P^i_{\;k} -\partial_j \Omega_\alpha P^j_{\;k}  + \partial_k\Omega_\alpha\right)^2\;.}\ee
{\it  Production is assumed to be an irreversible process and so, it can only take place iff  $ -\partial_i \Omega_\alpha P^i_{\;k} -\partial_j \Omega_\alpha P^j_{\;k}  + \partial_k\Omega_\alpha > 0$ and $ K^{kk}_\alpha >0$.}

\bigbreak
\noindent {\bf Proof: } 
The rate of change in satisfaction is given by,
\be\label{der} \frac{d}{dn^k}(\Delta_{{\cal M}} \Omega_\alpha) =   -\partial_i \Omega_\alpha P^i_{\;k} -\partial_j \Omega_\alpha P^j_{\;k}  + \partial_k \Omega_\alpha\;.\ee
In order for the agent to produce $k$, it must be able to increase it's satisfaction and so we need ${d(\Delta_{{\cal M}} \Omega_\alpha)}/{dn^k}  > 0$. As the agent keeps producing $k$ and depleting $i$ and $j$, it will make at least one of the derivatives $\partial_i \Omega_\alpha$ or $\partial_j\Omega_\alpha$ very large and positive. This follows from the convexity of the index of satisfaction. Thus there is a point where the derivative (\ref{der}) will change sign. This represents the equilibrium point.

Following the general spirit of a near-equilibrium expansion, we assume that the agent is close to that equilibrium point.  Integrating Eq. (\ref{der}) and expanding around $\Delta n^k  = 0$ we get,
\bea\label{dOMexp} \Delta_{\cal M} \Omega_\alpha &\approx&  \Delta n^k \left. \left( -\partial_i \Omega_\alpha P^i_{\;k} -\partial_j \Omega_\alpha P^j_{\;k}  + \partial_k\Omega_\alpha\right)\right|_{\Delta n^k = 0} \nonumber \\
&& + \frac{(\Delta n^k)^2}{2} \left. \left[ \partial_i^2 \Omega_\alpha (P^i_{\;k})^2 +\partial_j^2 \Omega_\alpha (P^j_{\;k})^2   + \partial_k^2 \Omega_\alpha + 2 \partial_i \partial_j \Omega_\alpha P^i_{\;k} P^j_{\;k} \right.\right.\nonumber \\
&& \left. \left.  - 2 \partial_i \partial_k \Omega_\alpha P^i_{\;k} - 2 \partial_j \partial_k \Omega_\alpha  P^j_{\;k}\right] \right|_{\Delta n^k = 0} + \ldots \eea
It is easy to see that the maximization equation,
\be \frac{\partial}{\partial \Delta n^k} \Delta_{\cal M} \Omega_\alpha = 0\;,\ee
gives Eq. (\ref{dnk}). Moreover, inserting (\ref{dnk}) in (\ref{dOMexp}) we get the change in satisfaction (\ref{dOM}).

\noindent$\square$

One can also consider a simpler process such as $i \rightarrow k$. The resulting equations can be obtained from the proposition above by setting $P^j_{\;k} = 0$.

\subsection{ Price Setting by a Market Maker}
Another economic operation that we will consider is the case where there is a special agent (Market Maker) which sets prices according to excess demand. This agent can interact at every time step with a different number of agents. This is different from the bartering process. In the later, two agents {\it negotiate} a price.  The Market Maker, however, satisfy the excess demand of his customers, but has the privilege of setting the stock prices. For simplicity, we will assume that the Market Maker is a wealth maximizer. One then can prove the following proposition:

\bigbreak
\noindent {\bf Proposition 4 :} {\it Let $M_n \equiv M^j_{\;i}$ the exchange rate between products $i$ and $j$ at time step $n$. Then, the Market Maker will adjust the exchange rate at every time step as, }
\be\label{prop4} M_n \approx M_{n-1}  +  \frac{1}{2}  \frac{ \sum_\alpha  J^{ii}_\alpha \left(\partial_i \Omega_\alpha - M_{n-1}  \partial_j \Omega_\alpha\right)}{\sum_\alpha  J^{ii}_\alpha   \partial_j \Omega_\alpha} \;,\ee
{\it where the sums go over all agents interacting with the Market Maker, and all quantities in the RHS are evaluated at time step $n-1$. }

\bigbreak
{\bf \noindent Proof:}  The excess demand for product $i$ from the buyers/sellers at price $M_n$ can be found from proposition 1,
\be D(M_n) = \sum_\alpha \Delta n^i_\alpha \approx \sum_\alpha  J^{ii}_\alpha \left(\partial_i \Omega_\alpha - M_n \partial_j \Omega_\alpha\right)\;.\ee
Therefore, the satisfaction of the Market Maker (MM) changes by,
\be \Delta \Omega_\text{MM} = \Omega_\text{MM}(n^i_\text{MM} - D(M_n), n^j_\text{MM}  + M_n D(M_n)) -  \Omega_\text{MM}(n^i_\text{MM} , n^j_\text{MM} ) \;.\ee
We assume that the excess demand is not big enough to deplete the MM inventory. This needs to be checked at every time steps in the simulations. 

The MM now maximizes his satisfaction w.r.t. $M_n$:
\be \frac{\partial}{\partial M_n} \Delta \Omega_\text{MM} = 0\;.\ee
This equation can be written as,
\be\label{MMmax} -\frac{\partial_i\Omega_\text{MM}}{\partial_j\Omega_\text{MM}} + \frac{D(M_n)}{D'(M_n)} + M_n = 0\;.\ee
Since the MM is a profit maximizer, his utility has the form,
\be U_\text{MM} \sim \log\left[ n^i_\text{MM} M_\text{MM}+ n^j_\text{MM} + \ldots \right]\;,\ee
where $M_\text{MM}$ is the ``model" that the MM has of the price.
Thus,
\be \frac{\partial_iU_\text{MM}}{\partial_jU_\text{MM}} = M_\text{MM}\;.\ee
We now assume that the MM expectation is such that $\text{E}[M_\text{MM}] = M_{n-1}$ where $M_{n-1}$ is the last closing price. Moreover, the MM predicts zero variance: $\text{Var}[M_\text{MM}] = 0$. Thus,
\be \frac{\partial_i\Omega_\text{MM}}{\partial_j\Omega_\text{MM}} = M_{n-1}\;.\ee
It is then easy to see that Eq. (\ref{MMmax}) reduces to Eq. (\ref{prop4}) of the proposition.

We can now easily check that the MM indeed makes a profit, regardless of the sign of the excess demand. The wealth of the MM at time $n-1$ is (measured in units of $j$), $W_{n-1}=( n^i_\text{MM} M_{n-1}+ n^j_\text{MM} + \ldots)$. At time $n$, we have  $W_{n}= ([n^i_\text{MM} - D(M_n)] M_{n-1}+ [n^j_\text{MM}  + M_n D(M_n)]+ \ldots)$. Thus, the change in wealth is,
\be \Delta W = W_n - W_{n-1} = D(M_n)(M_n - M_{n-1}) = 
\frac{[D(M_{n-1})]^2}{2 \sum_\alpha  J^{ii}_\alpha   \partial_j \Omega_\alpha} > 0\;.\ee

\noindent$\square$

This model of a Market Maker is very similar to other models in the literature \cite{he} which assume a linear relation between the price adjustment and excess demand. Here, however, we give a precise microeconomic derivation.

\subsection{What-by-What Matrices and Information}

In the previous  sections we have presented the four basic economic operators: an exchange at a fixed rate ($\hat{\cal T}_1$),  an exchange trough a barter process ($\hat{\cal T}_2$), a production process ($\hat{\cal M}$) and a centralized Market Maker who sets prices according to excess demand. We have given ``flow" equations describing how much stuff is traded/produced based on the maximization of satisfaction near equilibrium. Moreover, we have formulas for the change of satisfaction for fixed information. 

However, once a trade is performed, one has new information about the system: there is an exchange rate $-\Delta n^i/\Delta n^j$. How do agents process such information? Recall that agents carry models of the exchange rates, $(M_\alpha)^i_{\;j}$. A trading process produces a new data point which can be fed into such model. As we assumed in section 2, agents have enough rationality to maintain a {\it consisten} w-w matrix. Given a trade with only two products, say $i$ and $j$, there is no unique way of updating all entries of the w-w matrix. Here we propose a simple formula to do this. 

For the operation $\hat{\cal T}_1$ we propose that agents update their w-w matrix as\footnote{Note that this is an update of a new data point to the {\it model} of  $M_\alpha$.},
\be \label{consistency1} \hat{\cal T}_1:\;\;\;  (M_\alpha)^i_{\;j} \equiv \frac{\partial_j\Omega_\alpha}{\partial_i \Omega_\alpha} \;, \;\;\; \forall i,j \in {\cal P}\;,\ee
where we evaluate at the point $\Delta n^i = \Delta n^j = 0$ (that is, before the actual trade). 

For  the barter operation $\hat{\cal T}_2$ we propose a similar form,
\be\label{consistency2} \hat{\cal T}_2:\;\;\;  (M_\alpha)^i_{\;j}  = (M_\beta)^i_{\;j}\equiv \frac{\partial_j\Omega}{\partial_i \Omega} \;, \;\;\; \forall i,j \in {\cal P}\;,\ee
where we evaluate at the point $\Delta n^i = \Delta n^j = 0$ (that is, before the actual trade), and $\Omega = \Omega_\alpha + \Omega_\beta$. Note that in a barter both agents must agree on the final exchange rate and thus $(M_\alpha)^i_{\;j}  = (M_\beta)^i_{\;j}$. Finally we see that the definition (\ref{consistency2}) coincides with the actual exchange rate computed within the linear approximation, Eq. (\ref{dnjT2}). 
The conditions (\ref{consistency1}) and (\ref{consistency2}) can be relaxed in more general models.

\section{Simulations of Decentralized Markets}
In this section we consider some numerical simulations of simple decentralized market models. In these models, agents are paired at random. Such structure is known in the literature as ``search equilibrium" \cite{KW, search1, search2}. However, we emphasize that we are working out of equilibrium.  Once two agents find each other they trade trough a Barter operation ($\hat{\cal T}_2$). However, they can choose which trade to make such that they obtain the most mutual satisfaction according to Eq. (\ref{dOmegaT2}). In other words, they get to choose which pair of products to trade at every time step. Since every agent can be paired with any other, we see that the we have a {\it complete} network.

We start in subsection 5.1 with an example of a decentralized market with only two products and no production. We show how there is price discovery and explain the properties of the relaxation to equilibrium. Using a simple mean-field theory analysis we estimate the relaxation time. In section 5.2 and 5.3, we add the simple transformations $i \rightarrow j$ and $j\rightarrow i$. We find that the system can self-organizes into periodic price cycles with price convergence, or remain in a disordered steady-state where agents never converge to an equilibrium price.

In section 5.4  we address the interesting question of the emergence of (commodity) money. We consider the case of three products $i$, $j$ and $k$ but now with a simple transformation $i \rightarrow j$ and $j \rightarrow i$. In this case, $k$ is not being transformed and so it can only be traded. We find that for certain initial conditions, the system self-organizes into periodic price cycles. Moreover, the system can self-organize into a phase where all agents use $k$ as a currency. In other words, all trades involve either $i \leftrightarrow k$ or $j\leftrightarrow k$ but not $i \leftrightarrow j$. However we find that such phase is unstable. 

Using a mean field theory analysis, we give a purely statistical explanation for the emergence of money. We find that, quite generically, the product which becomes the currency must be distributed among the population with the greatest variance, and the smallest covariance with relation to the other products.

\subsection{Two Products and the Convergence to Equilibrium}
In this section we study the simplest economic model satisfying the properties of section 2. The model has $N$ agents and two products which we call $i$ and $j$. The utility has the form
\be\label{util} U_\alpha = \log\left[ (n^i_\alpha)^\nu+  [(M_\alpha)^i_{\;j}]^\nu (n^j_\alpha)^\nu \right]\;.\ee

We assume that the agents' expectations for the price for the next time step is the same price that they found in the previous time step. That is, $\text{E}[ (M_\alpha)^i_{\;j} |_{t+1}] = (M_\alpha)^i_{\;j} |_{t}$, where $(M_\alpha)^i_{\;j} |_{t}$ is the price found trough the bartering process in the previous time step. We further assume that the agents are completely certain about their prediction so that  $\text{Var}[(M_\alpha)^i_{\;j} |_{t+1}] = 0$. Therefore, it follows that $\Omega_\alpha = U_\alpha$. Let us now describe the algorithm.

\bigbreak
\noindent {\it \bf Model \#1}: 
\begin{enumerate}
\item Agents start with some (random) inventory $n^i_\alpha$ and $n^j_\alpha$ and price $(M_\alpha)^i_{\;j}$
\item At every time step, agents are paired at random (we take an even number of agents).
\item Every pair of agents $(\alpha,\beta)$  decides which trade ($\hat{\cal T}_2$) to perform according to which gives them the larger satisfaction in Eq. (\ref{dOmegaT2}). They perform their favourite trade and update their inventories as,
\be n^i_\alpha \rightarrow n^i_\alpha + \Delta n^i\;,\;\;\; n^j_\alpha \rightarrow n^j_\alpha + \Delta n^j\;,\;\;\; n^i_\beta \rightarrow n^i_\beta - \Delta n^i\;,\;\;\; n^j_\beta \rightarrow n^j_\beta - \Delta n^j\;,\ee
where $\Delta n^i$ and $\Delta n^j$ is given in Eqs. (\ref{dniT2}) and (\ref{dnjT2}) with the utility given in Eq. (\ref{util}) above. Note that all derivatives are taken before updating the inventory and the price.  They also update their prices using Eq. (\ref{consistency2}). 
\item The process is iterated from step 2.
\end{enumerate}

In figure 1 we show a typical result of such simulation. 
\myfig{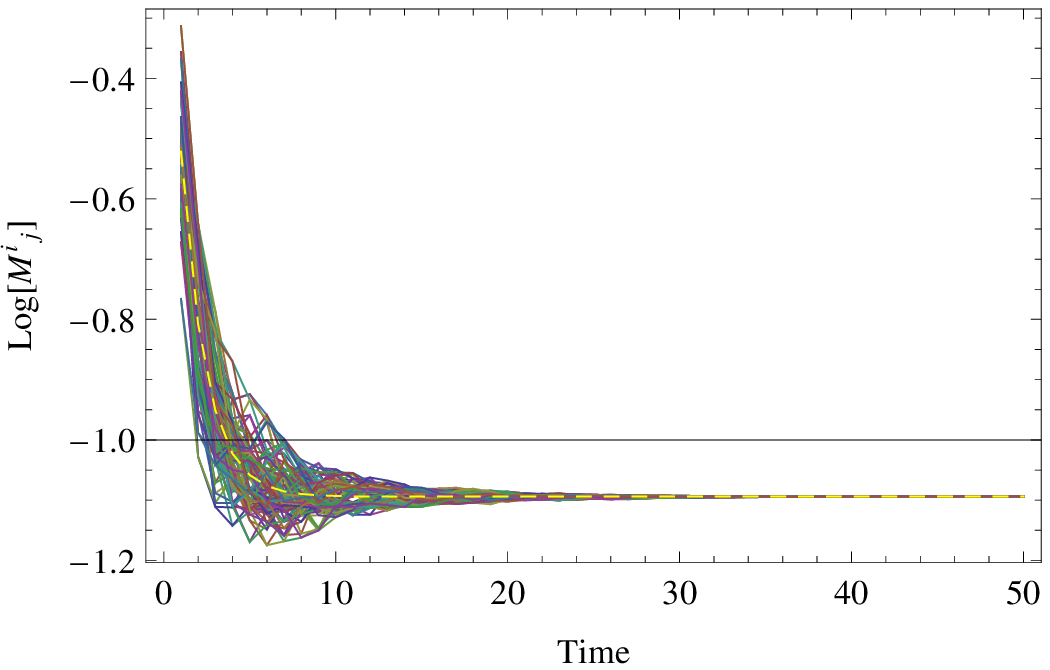}{10}{\small Relaxation to equilibrium with two products and no production. The yellow dashed line is the mean field theory result. } 
For the simulation we used $N = 100$ agents. We took the exponent of the utility to be $\nu = 1/2$. The initial inventories were taken from a uniform random distribution in the intervals: $n^i_\alpha \in [1,2]$, $n^j_\alpha \in [4,5]$. Note that the system is scale invariant and so the overall order of magnitude of the inventories is irrelevant. The initial prices were taken from a uniform random distribution in the interval $
(M_\alpha)^i_{\;j} \in [1,1.1]$. Figure 1 shows the prices $(M_\alpha)^i_{\;j}$ of every agent in different colors. We see the convergence towards a fundamental price. The horizontal line is the fundamental price which, by scale invariance and the form of the utilities, is simply $(M)^i_{\;j} = n^i/n^j$, where $n^i$ and $n^j$ are the total number of products $i$ and $j$ in the economy. 

We can get a better understanding of the relaxation to equilibrium by doing a mean field theory analysis of the pricing formula (\ref{consistency2}). Let $(M_\alpha)^i_{\;j} \equiv M_\alpha$ and let the brakets $\bra\cdot\ket$ denote averages over the population. Then we expand,
\be n^i_\alpha = \bra n^i\ket  + \delta n^i_\alpha \;,\;\;\; n^j_\alpha = \bra n^j\ket  + \delta n^j_\alpha \;,\;\; M_\alpha = \bra M\ket + \delta M_\alpha \;,\ee
for all $\alpha \in {\cal N}$.

To leading order Eq. (\ref{consistency2}) gives us the average price at time $t$ in terms of the price at $t-1$:
\be \bra M_t\ket \approx \bra M_{t-1}\ket^\nu \left(\frac{n^i}{n^j}\right)^{1-\nu}\;,\ee
Here we used the fact that since the product number is conserved the average product numbers are constant in time and that $\bra n^i\ket/\bra n^j\ket = n^i / n^j$.
One can easily iterate this equation to obtain,
\be\label{timeseries} \bra M_t\ket = \bra M_0\ket^{\nu^t}  \left(\frac{n^i}{n^j}\right)^{1 - \nu^t}\;.\ee
We see that as $t \rightarrow \infty$ the price $\bra M\ket$ goes to the fundamental one $n^i/n^j$. We can define the relaxation time for such process by $\nu^t \equiv e^{-t/\tau}$, and we find
\be \tau = - \frac{1}{\log \nu}\;.\ee
We see that for small $\nu$ the relaxation time is small and so equilibrium is reached quickly. For $\nu \rightarrow 1$ the relaxation time becomes infinite and the system takes longer to equilibrate. In fact, the price formula (\ref{timeseries}) fits the numerical data quite well as shown in figure 1.

Another view of the relaxation time can be obtained using the flow equation (\ref{dniT2}). Doing a similar mean field theory analysis we find, 
\bea \label{flow} \Delta n^i &\approx&  \frac{ \bra M \ket^\nu \bra n^j\ket^{\nu-1}}{2\left( \bra n^i\ket^\nu +\bra M\ket^\nu \bra n^j\ket^\nu\right)}\left( \bra n^i\ket \delta n^i_{\alpha \beta} - \bra n^j\ket \delta n^j_{\alpha \beta} \right) \nonumber \\
&&+  \frac{\nu  \bra M\ket^{\nu - 1} \bra n^i \ket \bra n^j \ket^\nu }{2(\nu - 1) \left( \bra n^i\ket^\nu +\bra M\ket^\nu \bra n^j\ket^\nu\right)} \delta M_{\alpha \beta}\;, \eea
where 
\be \delta n^i_{\alpha\beta} \equiv \delta n^i_\alpha - \delta n^i_\beta \;,\;\; \delta n^j_{\alpha\beta} \equiv \delta n^j_\alpha - \delta n^j_\beta \;,\;\;\; \delta M_{\alpha\beta} \equiv \delta M_\alpha - \delta M_\beta\;.\ee
The first term in Eq. (\ref{flow}) gives the linear response to gradients in product number. The second term is the linear response due to price gradients. We see that as $\nu \rightarrow 0$, the first term dominates and the system is very sensitive to inventory gradients. This is why the relaxation time is smaller. For $\nu \rightarrow 1$, the second term dominates. This means that agents are very sensitive to price movements, and hence are less sensitive to the distribution of products. We should also note that the limit $\nu \rightarrow 1$ is singular and so our linearized approximations break down.

A third way to look into the equilibrium state of the system is through the product distribution. The utility maximization implies,
\be d\Omega_\alpha = 0 \implies  \partial_i \Omega_\alpha - (M_\alpha)^i_{\;j} \partial_j \Omega_\alpha = 0\;. \ee 
Using the utility Eq. (\ref{util}) and the fact that $\Omega_\alpha = U_\alpha$ in this model, we get that in equilibrium, the inventory must be constrained such that
\be (M_\alpha)^i_{\;j} n^j_\alpha = n^i_\alpha\;.\ee
If the agents have roughly agreed on a price we get 
\be \bra M\ket  n^j_\alpha = n^i_\alpha\;.\ee
Therefore, we see that at equilibrium, agents' product numbers should lie in this line.
Figure 2 shows a scatter plot of agents inventories in the $(n^i,n^j)$ plane, and the equilibrium line $\bra M\ket n^j_\alpha - n^i_\alpha = 0$. The plot was taken at time step $t = 50$ for the simulation of fig. 1 ($\nu = 1/2$).
\myfig{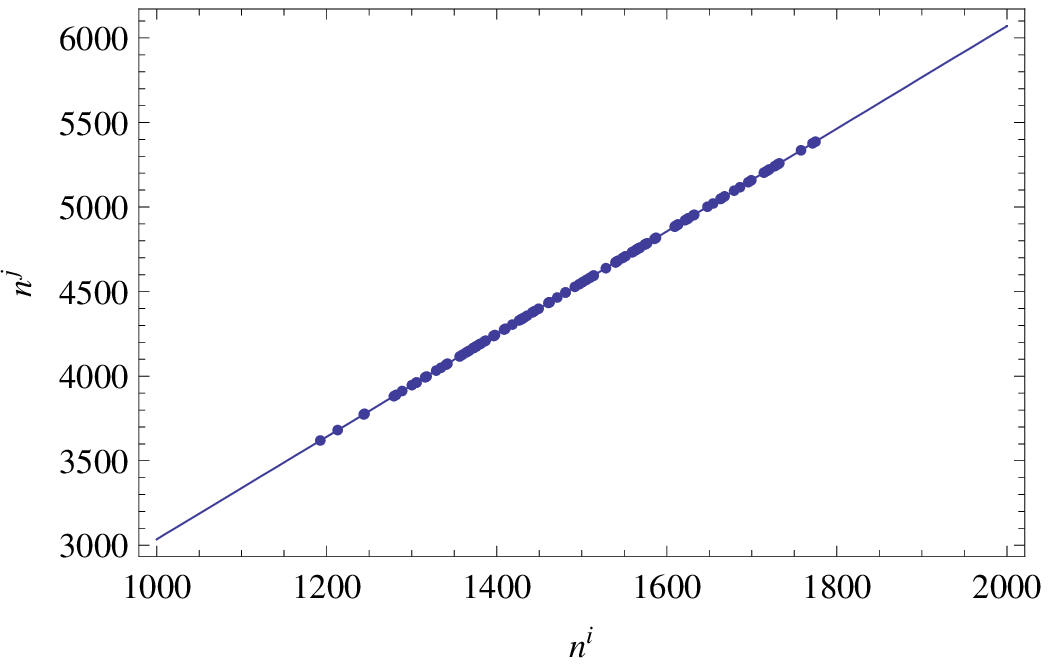}{10}{\small Equilibrium distribution of goods.} 
We see that indeed agents have reached equilibrium.

\subsection{Production, Price Cycles and Self-Organization}

We can now implement some production processes in our two product example. 
For this simple market there are only two possibilities: $i \rightarrow j$ and $j \rightarrow i$. We can specialize the results of proposition 3 to this simpler production operation. 
 For the process $i \rightarrow j$, we just take $P^i_{\;k} \equiv P^i_{\;j}$ and $P^j_{\;k}  = 0$. Similarly, for $j \rightarrow i$, we take $P^j_{\;k} \equiv P^j_{\;i}$ and $P^i_{\;k}  = 0$. 
The algorithm will be the same as the previous one, except that at every time step, we allow the agents to choose which production process to do before doing a trade.  That is, all agents have the same ``skills". They choose which process to perform according to which one gives them the best satisfaction, which can be computed from Eq. (\ref{dOM}).   Then they update their inventories according to Eqs. (\ref{dniM}) and (\ref{dnjM}) (with the appropriate indices). If both processes decrease their utility they choose not to produce.

We note that for $P^i_{\;j} > 1$ and $P^j_{\;i} > 1$, the output of each process will decrease the total number of products. Therefore, these operations will tend to decrease the utility and therefore will not be used by the agents. Now, for $P^i_{\;j} < 1$ and $P^j_{\;i} < 1$, the total number of products is increased. For example, for $P^i_{\;j} = 1/2$, we have a transformation $ 1j\rightarrow 2 i$.

In order to illustrate the self-organization of the system, we take a system with $N = 10$ agents and $\nu = 1/2$. Moreover, we take the initial inventory of all agents to be the same: $n^i_\alpha = n^j_\alpha = 1$ $\forall \alpha \in {\cal N}$. However, in order to start the dynamics we take the initial w-w matrix from a uniform distribution in the interval $(M_\alpha)^i_{\;j} \in [0.8,1.2]$.  Finally, the production rates are taken to be $P^i_{\;j}=P^j_{\;i} = 1/2$.

 In figure 3 we show the result of this simulation for the w-w matrix of all agents. We see that the system goes first trough a disordered phase but at some point it converges to a cyclical equilibrium state. The red dashed line is the fundamental price $M^i_{\;j} = n^i/n^j$. Note that due to the finite relaxation time, the agents are not completely converging to this price. Nevertheless, they do agree on a price between each other. One can show that the reason for the price cycles is that agents are twitching collectively between the two types of productions. 
\myfig{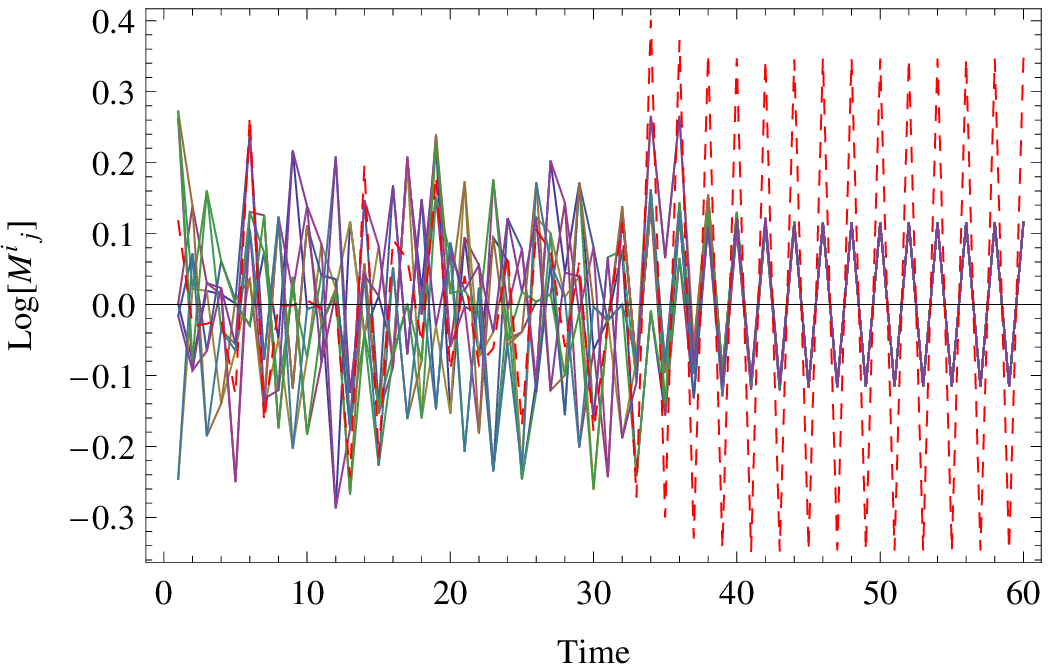}{10}{\small Self-organization of a two product system. The agents are performing the metabolic process $i\rightarrow j$ and $j\rightarrow i$. The red-dashed line is the equilibrium price $M^i_{\;j} = n^i/n^j$.} 
 
 Another important point is that the total number of products in the economy is growing exponentially in time. For example, in figure 4 we plot the log of the total number of product $i$ as a function of time.
 \myfig{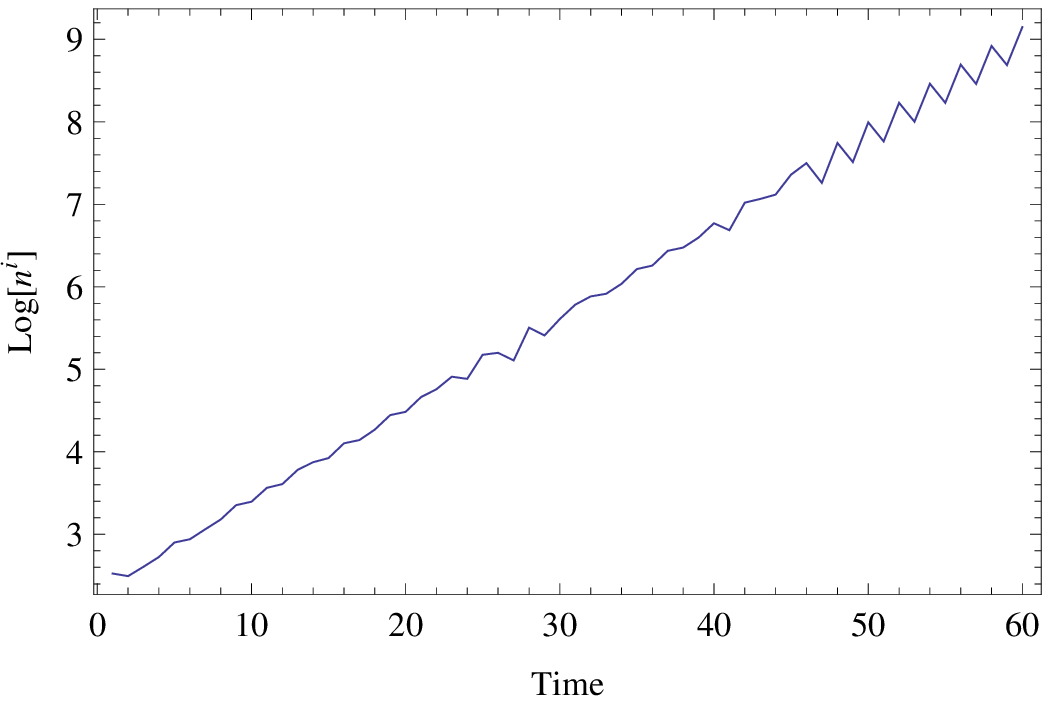}{10}{\small Log of the total number of product $i$ in the economy.} 

 \subsection{Not Equilibrium, but a Steady-State}
 So far we have seen cases where the agents collectively find an equilibrium price, even when it is time dependent. However, this is a rather special situation. More generally, the system never quite reaches an equilibrium. Rather, one can say that the system is in a {\it steady-state}. In such a state, the prices might not converge towards a unique equilibrium, but oscillate around it. Moreover, agents do not quite agree on the prices. In this section we show an example of such behavior. 
 
 The model is similar to the one of the previous section. However,  instead of letting the agents decide which production process they do, we assign them fixed skills. That is, they have the ability to do either $i \rightarrow  j$ or $j \rightarrow i$, but not both. 
 We distribute the skills randomly among the agents.  In figure 5 we show a typical run of such system. We have taken $N = 100$ agents, $\nu = 1/2$ and $P^i_j  = P^j_i = 0.9$. For the initial conditions we have, $n^i_\alpha = n^j_\alpha =  1$ and $(M_\alpha)^i_{\;j} \in [1,1.1]$. We note that the prices never quite reach an equilibrium.
  \myfig{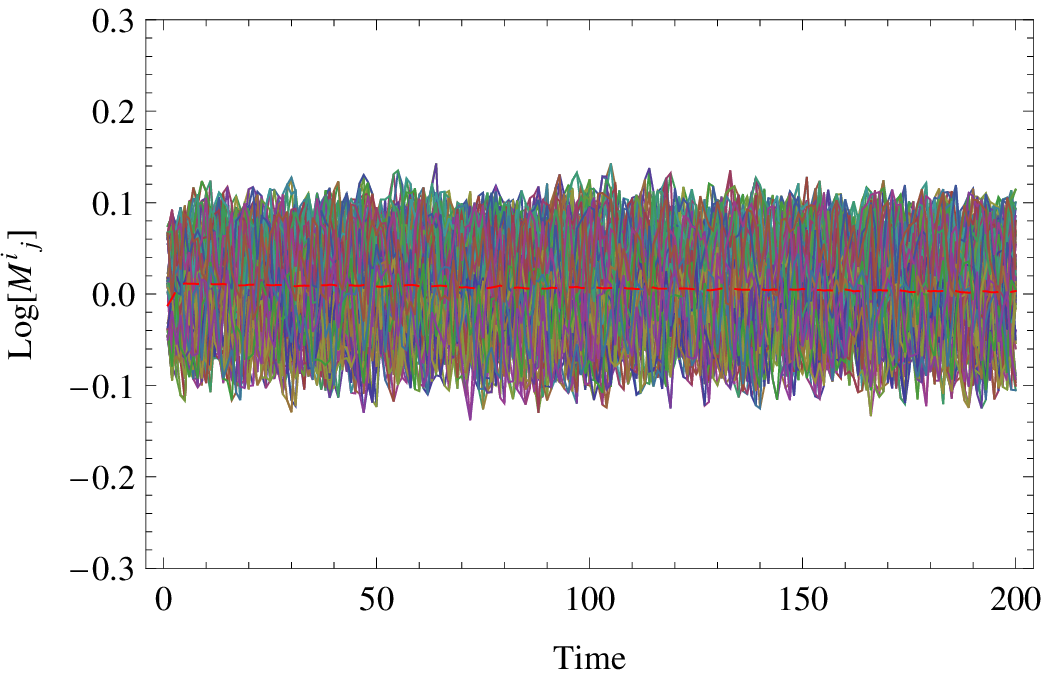}{10}{\small Disordered steady-state of a two product system. The agents are performing the metabolic process $i\rightarrow j$ and $j\rightarrow i$, but now with assigned skills. The red-dashed line is the equilibrium price $M^i_{\;j} = n^i/n^j$.} 

 In figure 6 we show the average of $\log M^i_{\;j}$, along with $\log(n^i/n^j)$, which is the ``equilibrium" price. We note that both are systematically away from each other. This shows that the system is truly in a non-equilibrium state. 
 \myfig{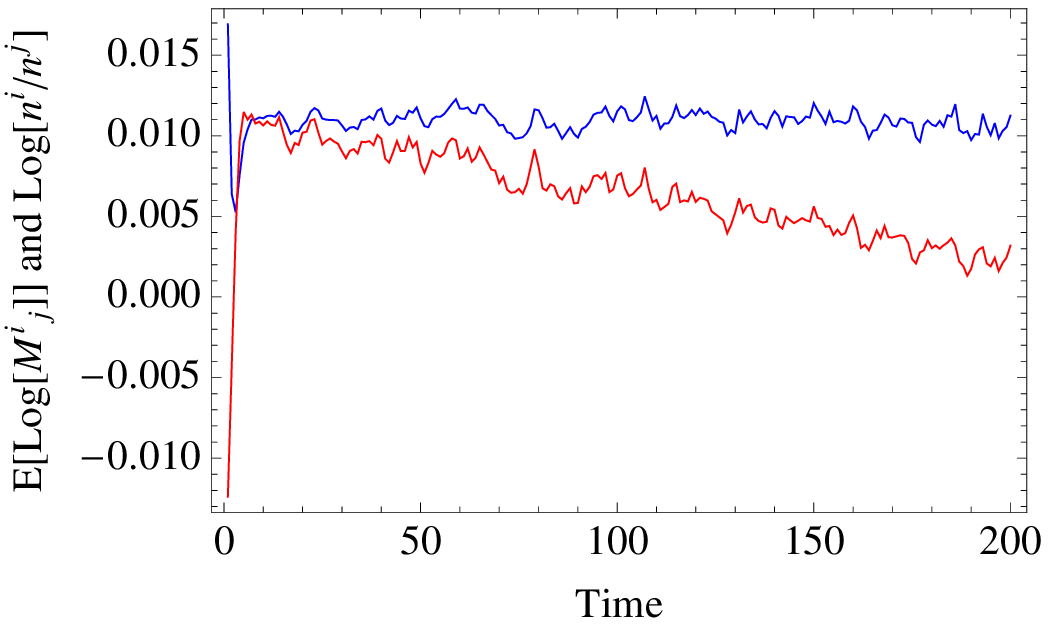}{10}{\small Comparison between the average log price $M^i_{\;j}$ (blue) and the ``fundamental one" $M^i_{\;j} = n^i/n^j$ (red) for the disordered steady-state of figure 5.} 

\subsection{The Emergence of Money}
In this section we study the emergence of (commodity) money in a simple system with three product and production. We want to emphasize that the ``currency" that we will see emerging correspond to commodity money as it is an object that give agents satisfaction. This should be contrasted with ``fiat" money which is intrinsically worthless to the agents. There is a large body of literature regarding the emergence of money.  In the classic paper \cite{KW}, Kiyotaki and Wright (KW) study 
a simple general equilibrium model with three goods and $N$ agents that could produce one of three kind of products and consume one of them (which is different from the one they produce). The agents find each other at random and have very simple strategies of ``trade or not-trade". The model also included the cost of storing for the different products. The authors found Nash equilibria where either commodity or fiat money serving as the medium of exchange. This kind of general equilibrium models where agents find each other at random is usually known as ``search equilibria". There have been many subsequent works following the KW model, see e.g. \cite{search1,search2}. For an interesting previous attempt to explain the emergence of money see \cite{jones}. 

Even though these models are interesting in that they show the existence of Nash equilibria with commodity or fiat money as a medium of exchange, they do not provide a explanation on how such transition could take place from a primitive barter economy. An attempt to explain the transition from a barter to a fiat money economy was developed in \cite{ritter}. The transition is explained still in the framework of general equilibrium theory by introducing a centralized government agent which ``must be able to promise credibly to limit the issue of money". This approach has been criticized by Selgin \cite{selgin} which claims that there is no known direct transition from barter to fiat money in history. Instead, it seems that fiat money is a rather recent development and that it has always followed an existing commodity money standard (or that the fiat money was convertible to some commodity such as gold).

In order to see a dynamical transition from barter to commodity money, one needs to go beyond equilibrium theory. The author is not aware of any non-equilibrium model that shows this transition. However, we should mention the models of \cite{Bak1,Bak2} which show that the price of fiat money can be determined dynamically. One important reason to have a non-equilibrium theory with a monetary phase transition, is to address the question of monetary stability \cite{stability1, stability2}. 

 In fact, we will find that in our first example the monetary phase is unstable. Before describing the simulations, we want to make a mean field theory analysis of the bartering process described in proposition 2. Suppose two agents find each other, and must decide which products to trade. In the near-equilibrium approximation, both agents will gain the same satisfaction as we showed in proposition 2. Thus, they will be able to unambiguously decide which pair to trade according to Eq. (\ref{dOmegaT2}). Let us now suppose that all agents have reached, more or less, an agreement on the exchange rate of products. In fact, they can disagree on the prices, but we will need to assume that such disagreement is much less than the inventory fluctuations. In equations,
 \be \frac{\delta M}{\bra M\ket}  \ll \frac{\delta n}{\bra n\ket }\;,\ee
 where the brackets indicate averages over the population, and $\delta n = n - \bra n \ket$, etc. are the fluctuations.

  We can now expand the gain in satisfaction, Eq. (\ref{dOmegaT2}), for a trade of $i$ and $j$ between agents $\alpha$ and $\beta$ around the averages. To leading order, we get  
 \be\label{dOmegaij} \Delta \Omega_{ij} \approx  \frac{\nu (1-\nu)\left(\bra n^i\ket \bra n^j\ket M^i_{\;j}  \right)^\nu}{8\left(\bra n^i\ket^\nu + (M^i_{\;j} \bra n^j\ket)^\nu \right) \sum_k (M^i_{\;k} \bra n^k\ket )^\nu} \left(\frac{\delta n^i_\alpha}{\bra n^i\ket} - \frac{\delta n^j_\alpha}{\bra n^j\ket}  - \frac{\delta n^i_\beta}{\bra n^i\ket} + \frac{\delta n^j_\beta}{\bra n^j\ket}\right)^2  \;. \ee
 We remind the reader that we are assuming utility functions of the form (\ref{util1}).

 We see that the terms with the averages on the RHS of Eq. (\ref{dOmegaij}) will not vary much according to which product the agents trade. The terms  in the parenthesis are the ones which give the largest fluctuations. These last terms will determine what products the two agents will decide to trade. We can now take an average over the whole population and get,
  \be \label{avdOmega}\bra \Delta \Omega_{ij} \ket \sim \bra (x^i - x^j)^2 \ket = \text{Var}(x^i) +\text{Var}(x^j) - 2 \text{Cov}(x^i,x^j)\;.\ee
 where we have defined the variables
 \be x^i = \frac{n^i - \bra n^i\ket}{\bra n^i\ket}\;.\ee
 
 For a product to become the favorite medium of exchange, it will need to maximize (\ref{avdOmega}). This can happen in various ways. From Eq. (\ref{avdOmega}) we see that such product (call it $i'$) must have the greatest variance ($\text{Var}(x^{i'}) \gg \text{Var}(x^i), \forall i \neq i'$) and/or the smallest covariance with the rest of the products ($\text{Cov}(x^{i'},x^i) \approx 0, \forall i\neq i'$). This last condition means that such product should be distributed pretty much independently from the other products.  Now let us see how can this happen in some numerical simulations.

Lets begin by studying a barter economy with three products $i$, $j$ and $k$ and the transformations the transformations $i \rightarrow j$ and $j\rightarrow i$.
That is, the model is just like the one we studied in section 5.2, but now with an extra product $k$ that does not gets transformed.
 So what would be the signature of a monetary phase? We say that the system is in a monetary phase with currency $k$ if all (or most) agents make the trades $i \leftrightarrow k$ and $ j\leftrightarrow k$ but not $i\leftrightarrow j$. Note that the existence of this phase must be due to the dynamics of the system as the agent utilities are invariant under changes of inventory measure. In other words, all products are treated in equal footing in a trading process. 
 
 We have found that, in this model, the monetary phase only exist for some relatively fine tuned initial conditions. As an example, we take $N = 100$ agents with $\nu = 0.1$. We also take the rate of production to be close to one: $P^i_{\;j} = P^j_{\;i} = 0.95$. This means that the total number of $i$ and $j$ will grow slowly. Initially, we give every agent the same amount of stuff: $n^i_\alpha = n^j_\alpha = 1$.   Finally, we take our initial w-w matrix from a uniform random distribution in the interval $(M_\alpha)^i_{\;j} \in [1,1.01]$ and $(M_\alpha)^i_{\;k} \in [1,1.01]$. Figure 7 shows the $(M_\alpha)^i_{\;j}$  entry of the w-w matrix for the first 20 time steps. All agents' are plotted in the same figure, but their agreement is so good that they are hard to distinguish. The dashed red line is the fundamental price $M^i_{\;j} = n^i/n^j$. 
\myfig{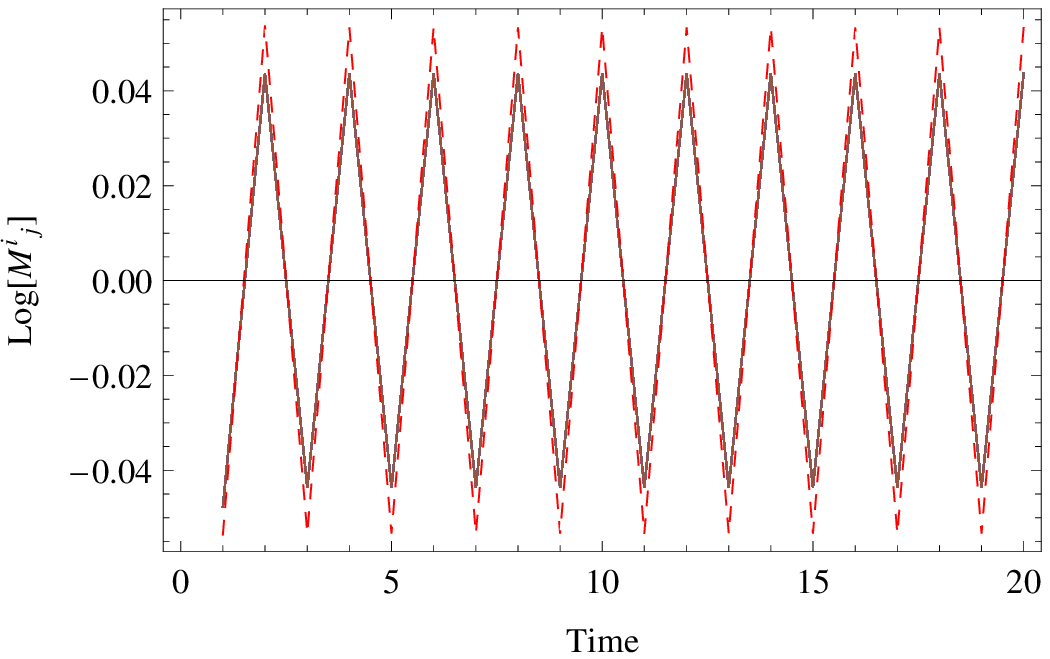}{10}{\small Log of the relative exchange rate of $i$ and $j$ for an economy with three products $i$, $j$, $k$, and the production processes $i \rightarrow j$ and $j \rightarrow i$. } 

The price cycles are what we expected from the previous simulations.  Now comes the surprising part. In figure 8 we plot the fraction of agents using product $k$ in a trade.  We see that for the first $~60$ time steps, more than 95\% of all agents are using $k$ in a trade. This is the monetary phase! However, what is surprising that after $t \sim 70$, the agents cease to use $k$ as a currency. That is, they are bartering with all possible combinations. We have plotted the line 2/3 since this is the fraction of agents we expect to use $k$ in a trade if the trades are being chosen at random. We don't observe any particular movement in the prices that coincides with the end of the monetary phase.
\myfig{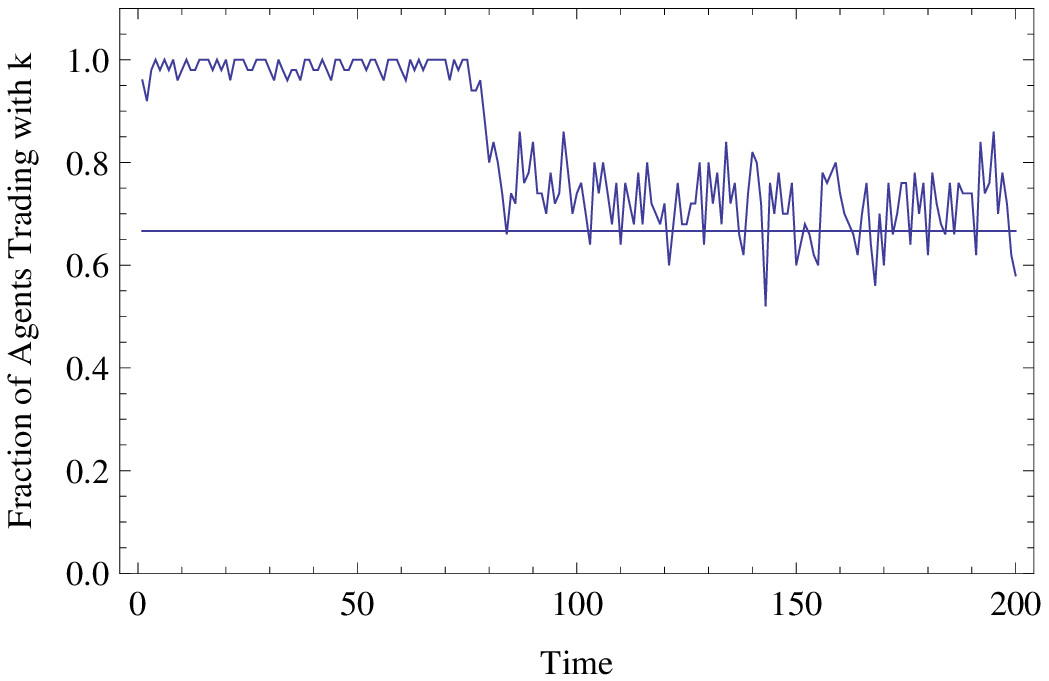}{10}{\small Fraction of agents using $k$ in a trade. The monetary phase is the plateau around the interval $t \in[0,60]$. Note that during this phase more than 95\% of the agents are using $k$ in a trade.} 

We now want to understand the existence and stability of the monetary phase using the mean field theory analysis described above. First, we note that ignoring the price discrepancies between agents is a very good approximation in this case. In figure 9 we plot  $\log\bra (x^i  - x^j)^2\ket$ for the different pairs of products. In the plot, we have only shown a time interval around the end of the currency phase, to make it more clear. First, we can see that the monetary phase is not an equilibrium as the variances are changing over time. Furthermore, we see that during this phase $\bra (x^i  - x^k)^2\ket  \approx\bra (x^j  - x^k)^2\ket >\bra (x^i  - x^j)^2\ket$. This is what we expect from the mean field theory analysis. However, once the product variances stabilize (a true ``equilibrium"), we find that  $\bra (x^i  - x^k)^2\ket  \approx \bra (x^j  - x^k)^2\ket \approx\bra (x^i  - x^j)^2\ket$, and so no trade pair is preferred over the others. This is the end of the monetary phase.
\myfig{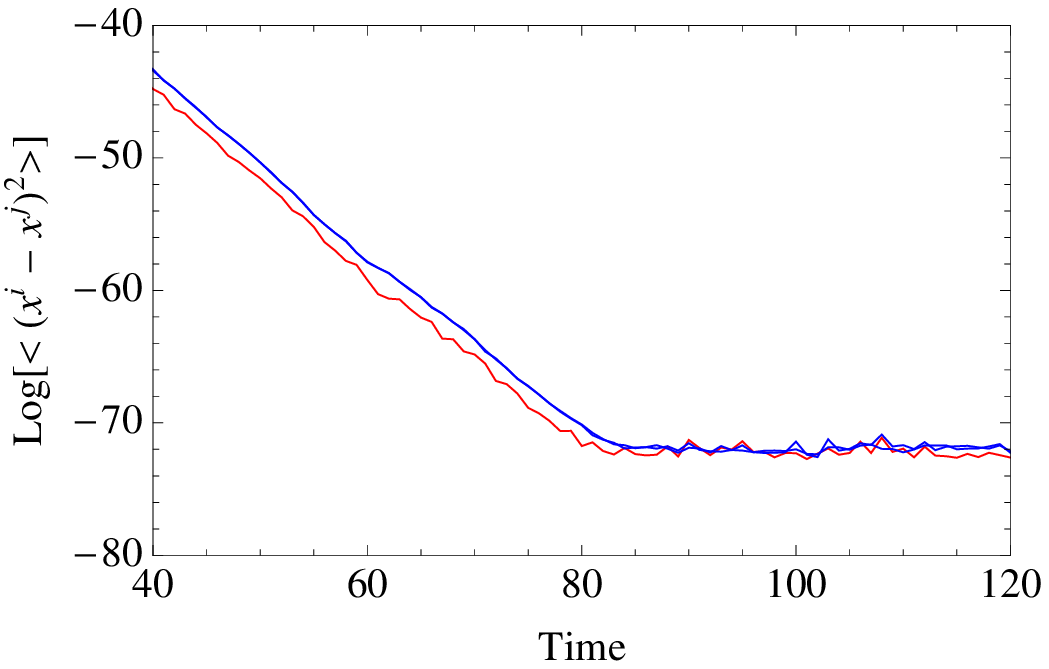}{10}{\small Log of the averages $\bra (x^i  - x^j)^2\ket$ for each pair of products. The blue lines is for $\bra (x^i  - x^k)^2\ket$ and $\bra (x^j  - x^k)^2\ket$. The red line is for $\bra (x^i  - x^j)^2\ket$. We see that $\bra (x^i  - x^k)^2\ket  \approx\bra (x^j  - x^k)^2\ket >\bra (x^i  - x^j)^2\ket$ in the monetary phase. So according to Eq. (\ref{avdOmega}), agents prefer (on averge) to trade with $k$. The end of the monetary phase is the plateau after $t \sim 80$. }  

The example above is interesting in that the monetary phase arises ``spontaneously" from the intrinsic dynamics of the system. However, this is not the only example where agents choose a preferred medium of exchange. In fact, according to the mean field theory analysis, all we need to have is a product with larger variance and smaller covariance than the rest. We can make this happen ``artificially" by endowing agents with a particular product at every time step. As an example, we study the same model as above, but without production. However, we assume that the amount of product $k$ that different agents have varies according to,
\be n^k_\alpha(t) = n^k_\alpha(t-1)(1 + r)\;,\ee
where $t$ is the time step, and we take $r$ from a uniform distribution with zero mean\footnote{Note that we take the return $r$ from the same distribution at every time step.}. Therefore, the total amount of product $k$ will not change on average. 

In figure  10 we show the fraction of agents bartering with product $k$. We see the roughly 90\% use $k$ in an exchange. In figure 11 we plot  $\log\bra (x^i  - x^j)^2\ket$ for the different pairs of products. We see that indeed the trades $i \leftrightarrow k$ and $j \leftrightarrow  k$ are preferred over $ i \leftrightarrow j$. This makes $k$ the favorite medium of exchange.  In this simulation, the agents do not completely agree on the prices. In figure 12 we plot the variances of the $M^i_{\;j}$ and $M^i_{\;k}$ entry of the w-w matrix. We see that they do not go to zero. This is another example of a steady-state rather than an equilibrium.

\myfig{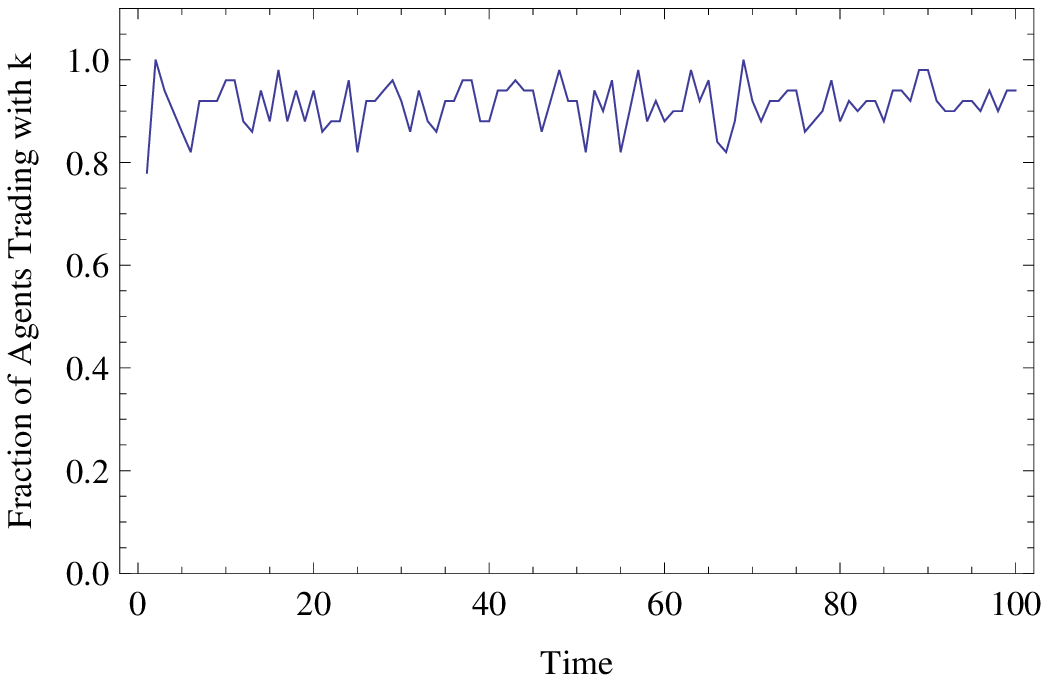}{10}{\small Fraction of agents using $k$ in a trade. This is with no production, but with an exogenous endowment of $k$ as described on the text.} 

\myfig{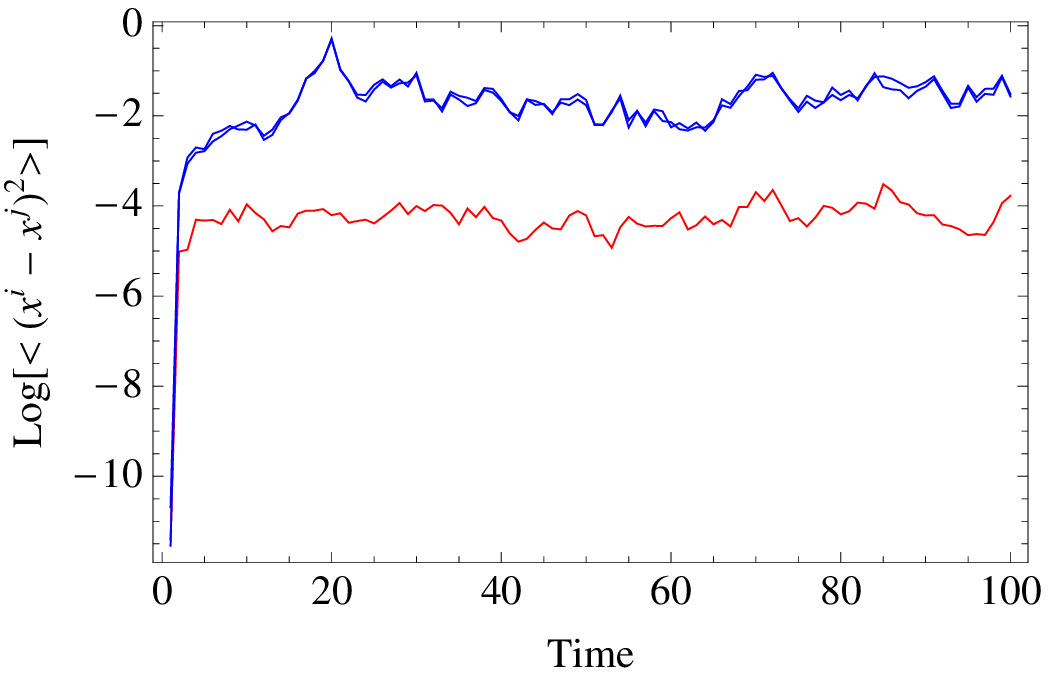}{10}{\small  Log of the averages $\bra (x^i  - x^j)^2\ket$ for each pair of products. The blue lines is for $\bra (x^i  - x^k)^2\ket$ and $\bra (x^j  - x^k)^2\ket$. The red line is for $\bra (x^i  - x^j)^2\ket$. } 

\myfig{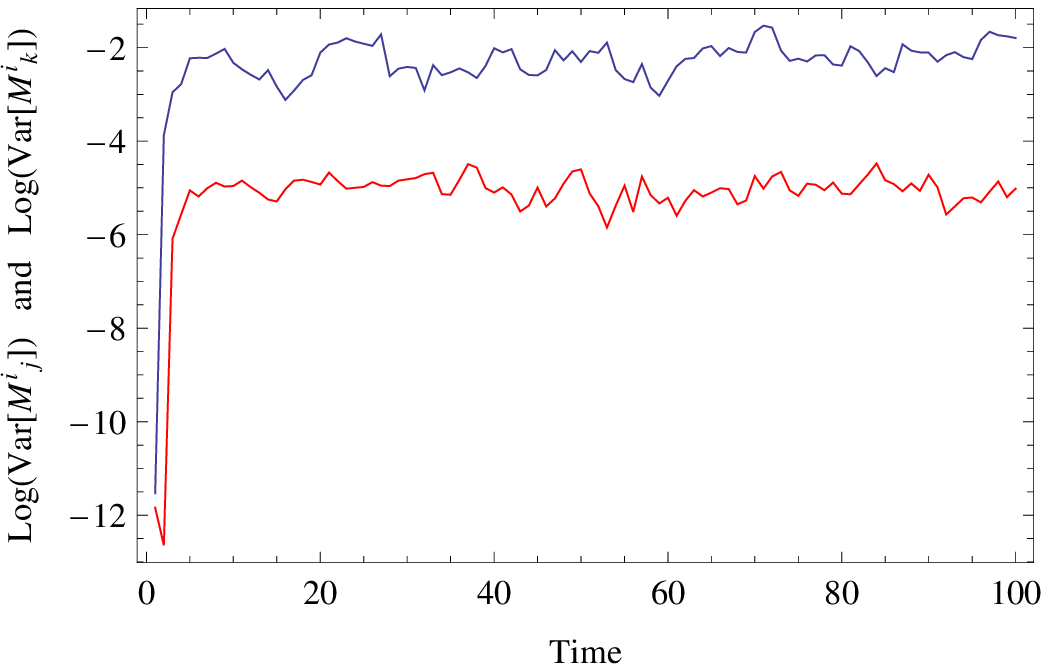}{10}{\small Log of variances of the w-w matrices. The red line is  $\log(\text{Var}[M^i_{\;j}])$ and the blue is $\log(\text{Var}[M^i_{\;k}])$.} 

\section{Simulations of a Centralized Market}
In proposition 4, we defined a centralized market with a Market Maker which sets prices according to excess demand. This structure is more appropriate in order to simulate financial markets. There is a large body of work on  agent based simulations of financial markets (see the review \cite{agentreview} and references therein). One important line of research along these lines is the impact of agents' heterogeneity to stock price movements (see the recent papers \cite{coulon,he} and the reviews \cite{hommes,lebaron}). In fact, it has been claimed that some of the most salient stylized facts of financial time series such as, fat tails, volatility clustering, and decaying autocorrelations are, in fact, due to the heterogeneous strategies and expectations of the market participants.  

Our purpose in this section is not to provide new insights into these phenomena. Instead, we want to illustrate that the framework presented in this paper is very appropriate to study the effects of agents' heterogeneity. 

 To this end we present some simple simulations of centralized markets. We begin in section 6.1 with the simplest model of homogeneous expectations, in order to see the expected convergence to equilibrium. In section 6.2  we consider a model where agents make speculations about the price return of the next time step. We show the emergence of a speculative bubble and a subsequent crash. In all simulations we will use only two products for simplicity. Moreover the utility will be of the usual form, Eq. (\ref{util}). In the following we will use the notation $M \equiv M^j_{\;i}$. The basic algorithm for all the models is as follows,

\bigbreak
\noindent {\it \bf  Model \# 2:}
\begin{enumerate}
\item Agents are started with some random initial inventories and expectations on the next w-w matrix value. This step will be explained in more detail below.
\item A new market closing price is computed from Eq. (\ref{prop4}). We use a market maker with an infinite supply of products and so we do not have to worry about depleting his inventory.
\item Agents update their inventories using 
\be n^i_\alpha \rightarrow n^i_\alpha + \Delta n^i_\alpha\;,\;\;\; n^j_\alpha \rightarrow n^j_\alpha - M_n \Delta n^i_\alpha\;,\ee
where,
\be \Delta n^i_\alpha = J^{ii}_\alpha \left(\partial_i \Omega_\alpha - M_n \partial_j \Omega_\alpha\right)\;,\ee
and $M_n$ is the closing price.
\item Agents update their beliefs for the next time step.
\item The process is repeated from step 2.
\end{enumerate}

\subsection{Homogeneous Expectations}
In the simplest model, agents will have homogeneous expectations about the next w-w matrix value. First, consider a fixed expectation,
\be \text{E}[(M_\alpha)_n] = M_{n-1}\;.\ee
That is, agents agents believe that the next price is the same as the previous one. This is the model of {\it rational expectations}. Furthermore, we can assume for simplicity that agents are certain about this prediction and so $\text{Var}[(M_\alpha)_n] = 0$.  Therefore, their satisfaction indices will be of the form,
\be \Omega_\alpha =  \text{E}[U_\alpha] = \log\left[(n^i_\alpha M_{n-1})^\nu  + (n^j_\alpha)^\nu \right] \;.\ee

In figure 13 we show the result of a simulation where agents have such and homogeneous beliefs. We took $N = 100$ agents with $\nu = 0.5$, and the initial endowments where taken from a uniform distribution in the interval $n^i_\alpha \in [1,5]$ and $n^j_\alpha \in [1,8]$. The initial closing price was taken to be $M_0 = 1$. We see that, as expected, agents reach an equilibrium which, by scale invariance, is $M_\text{eq} = n^j/n^i$. Note that the total number of products is not conserved here since we assumed a market maker with an infinite supply.
\myfig{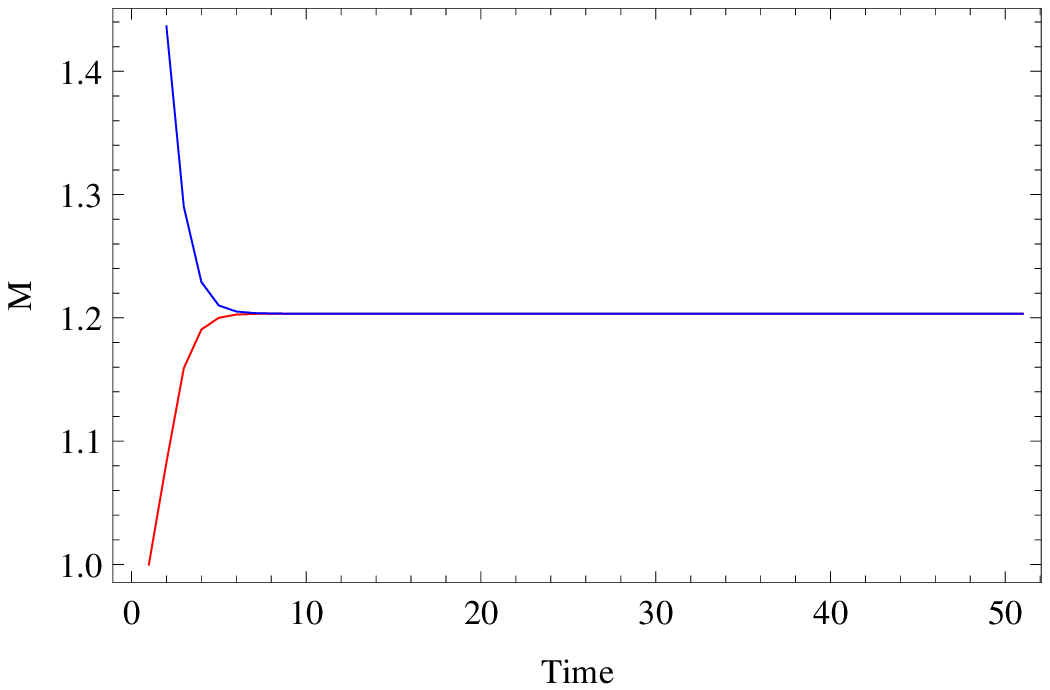}{10}{\small Relaxation to the equilibrium price in a centralized market with homogeneous expectations.} 

We note  that the w-w matrix is an internal valuation tool, and so it can be interpreted as a ``fundamental" price which comes from some exogenous information that the agent might have at hand. We should note that agents which hold a fixed belief about a stock price are usually called ``fundamentalist" in the literature \cite{he}. This is in order to distinguish them from the ``chartist" which make a prediction based on a previous price trend. Therefore, the model that we are considering here can be interpreted as consisting of fundamentalist only. A popular way of modeling fundamentalist is to assume that they hold {\it fixed} beliefs about the price of a stock \cite{he}. For example, our agents can have expectations based on some internal model such as the log-normal random walk:
\be\label{model} \text{E}[(M_\alpha)_n] = (M_\alpha)_{n-1}\;,\;\;\; (M_\alpha)_{n} =  (M_\alpha)_{n-1} (1 + r_\alpha) \;,\ee
where the return $r_\alpha$ is taken from a gaussian normal distribution with zero mean (at every time step). The index of satisfaction can now be approximated, to leading order in the variance of $r_\alpha$ as,
\bea \Omega_\alpha &\approx& \log\left[(n^i_\alpha (M_\alpha)_{n-1} )^\nu  + (n^j_\alpha)^\nu \right] \nonumber \\
&& -\frac{ \nu (n^i_\alpha (M_\alpha)_{n-1})^\nu
   \left[(n^i_\alpha (M_\alpha)_{n-1} )^\nu+ (1 - \nu)(n^j_\alpha)^\nu\right] }{2 
  \left[(n^i_\alpha (M_\alpha)_{n-1} )^\nu  + (n^j_\alpha)^\nu \right]^2} \sigma_\alpha^2 + \ldots\nonumber \\
   \eea
where $\sigma_\alpha^2 = \text{Var}[r_\alpha]$. It is usually assumed that the model (\ref{model}) comes from some fundamental information about the ``company", which all agents have access to. Therefore, all agents will have the same model. 

In this case, one expects the market closing prices $M_n$ to follow closely the agents' model (\ref{model}). However, we need to remember that our agents might put more importance to the actual distribution of products than to their w-w matrix. The parameter that controls the relative importance of their w-w matrix and their inventories is the exponent $\nu$ in the utility.  For small $\nu$ the agents will put more importance to the ``fundamental" price $M = n^j/n^i$, than to their models. 
This will be important in triggering the market bubble in the next section.

Here we illustrate this by running a simulation with $N = 10$ agent, where all agents have the same model (\ref{model}) and they all have the same prediction for the return at every time step. We take $\nu = 0.5$ as an illustration. The initial distribution of goods is taken from the same uniform distribution as before. Moreover, we take the variance of the model (\ref{model}) to be $\sigma_\alpha = 0.01$. 
  In figure 14 we show the closing prices along with the model prices and the ``fundamental" equilibrium prices $M  = n^j/n^i$. We see that the actual closing prices are half way between the model and the fundamental ones, as expected.  If we take $\nu = 0.9$ instead, we see in figure 15 that the closing prices reasemble more the model price. 
If we want to see fat tails, volatility clustering, etc. we need to add some chartist strategies as in \cite{he}. We won't pursue this here. 
\myfig{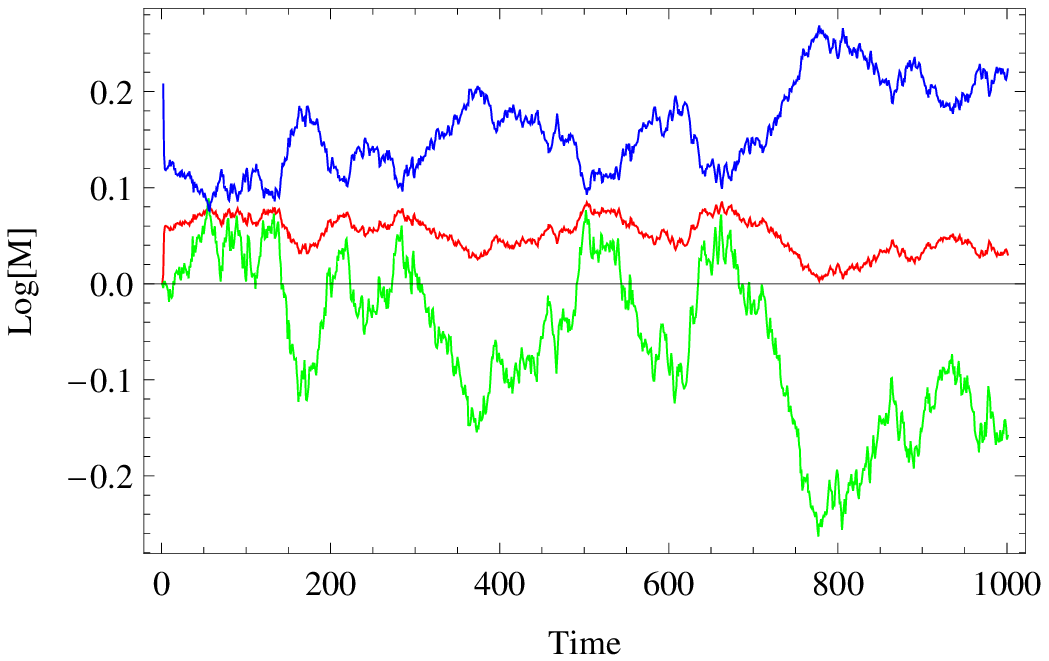}{10}{\small Closing prices for a market where all agents are ``fundamentalist" whose internal price is given by the random walk of Eq. (\ref{model}). The red line is closing price. The blue line is fundamental equilibrium price $M = n^j/n^i$, and green the agents' model. This is for a model with $\nu = 0.5$.} 

\myfig{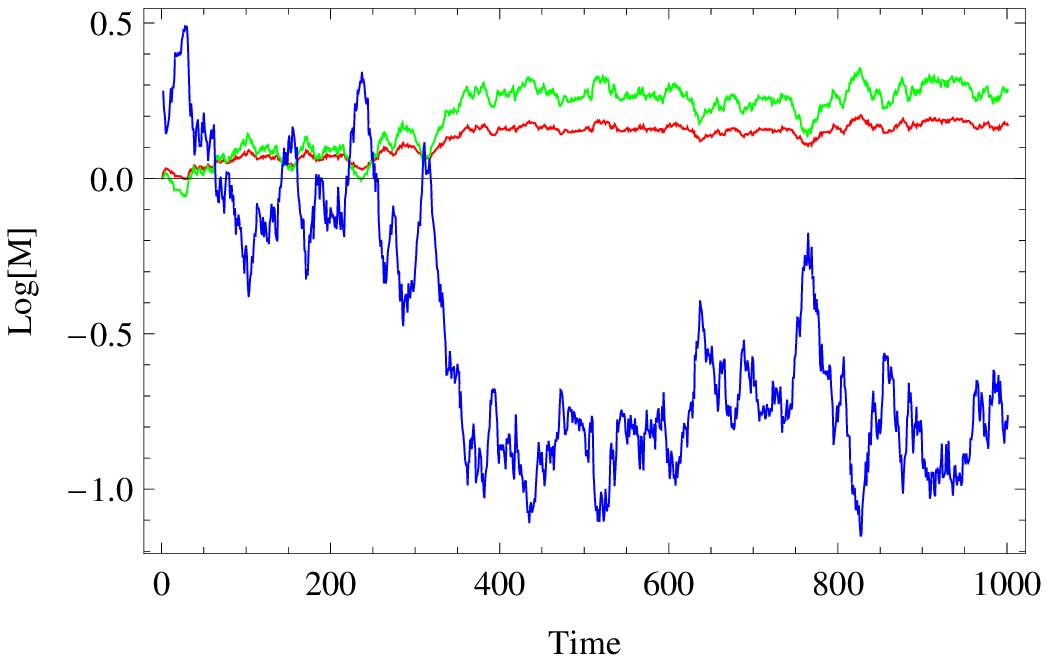}{10}{\small Closing prices for a market where all agents are ``fundamentalist" whose internal price is given by the random walk of Eq. (\ref{model}). The red line is closing price. The blue line is fundamental equilibrium price $M = n^j/n^i$, and green the agents' model.  This is for a model with $\nu = 0.9$.} 

\subsection{A Speculative Bubble}
In this section we present a simple example of a speculative bubble (see e.g. \cite{giardina}). As we have seen the parameter $\nu$ controls how much weight an agent puts in his own model of the market, versus the ``fundamental" price which comes from the total number of goods in the economy. In the first model that we presented in the previous section the agents had the simplest prediction $E[(M_\alpha)_n] = M_{n-1}$. Thus, they took information from the market and incorporated it in their models. In the second example, they had fixed beliefs, $E[(M_\alpha)_n] = (M_\alpha)_{n-1}$, and so they where totally ignoring the market's closing prices.

Here we want to construct a model that is in between these two. The agents' prediction will be of the form,
\be E[(M_\alpha)_n] = M_{n-1} e^{r_\alpha}\;,\ee
where now $r_\alpha$ is a {\it speculation} on what the log return should be in the next time step. We let all agents choose their own values of $r_\alpha$ at every time step. For simplicity we assume that all agents are certain about their beliefs and so $\text{Var}[(M_\alpha)_n] = 0$, and that all agents' beliefs $r_\alpha$ are taken from the same gaussian distribution. 

The index of satisfaction is then,
\be \Omega_\alpha =  \text{E}[U_\alpha] = \log\left[(n^i_\alpha M_{n-1}e^{r_\alpha})^\nu  + (n^j_\alpha)^\nu \right] \;.\ee
In figure 16 we show the results of a simulation with $N = 500$ agents, $\nu = 0.5$, and the initial inventory endowments are taken to be $n^i_\alpha  = n^j_\alpha = 1$. We also take the returns $r_\alpha$ from a normal gaussian distribution with mean $\mu = 4$ and standard deviation $\sigma = 2$. At every time step we take $r_\alpha$ from that distribution. Thus we see that agents are consistently betting on a positive return (on average). 

Figure 16 shows that at the beginning, the agents' speculation drive the market prices up. However, the fundamental price is falling so fast that the market price cannot keep with it. Eventually the bubble bursts and the market price is driven to zero. 
\myfig{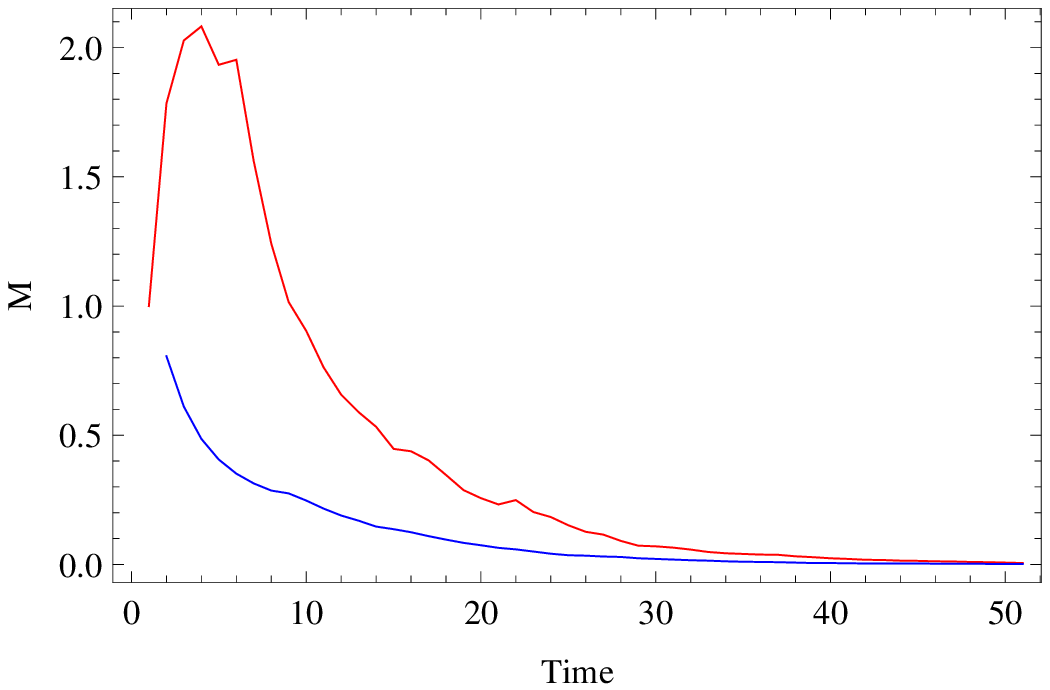}{10}{\small A speculative bubble. The red line is the actual market closing price. The blue curve is the ``fundamental" price $M^j_{\;i} = n^j/n^i$. This simulation is for $\nu = 0.5$.} 

The existence of the bubble lies in the fact that agent pay attention to {\it both} their inventories and their beliefs. If we take a larger value of $\nu$, say $\nu = 0.9$, agents will put more weight to their beliefs than to their inventories, and they can even turn around the fundamental price as shown in figure 17. 
\myfig{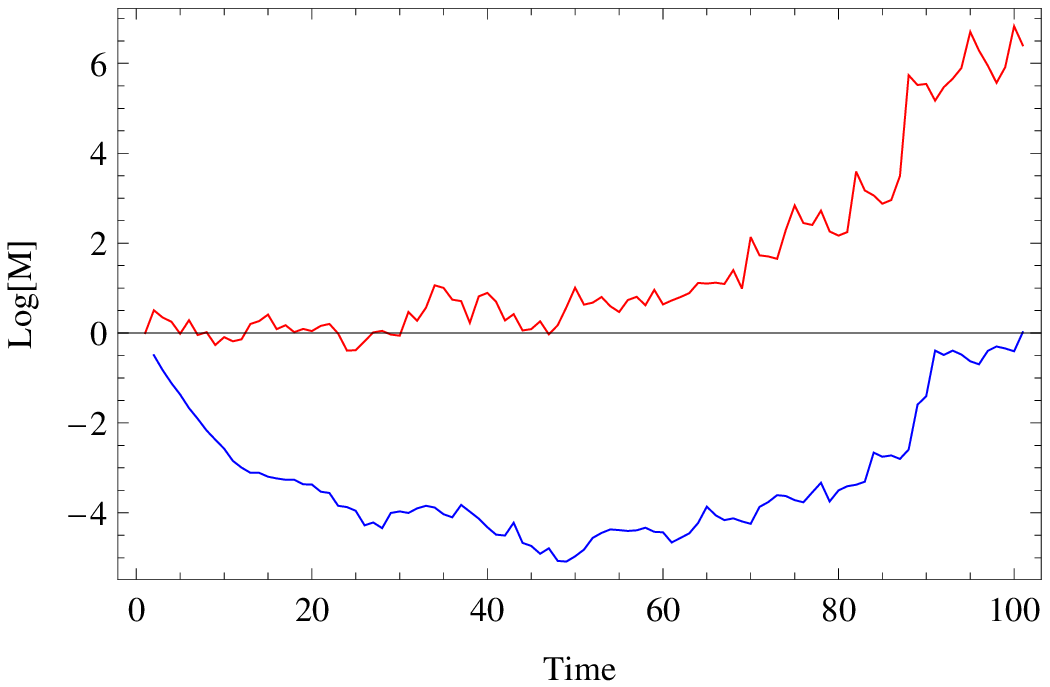}{10}{\small Agent's beliefs can turn around the fundamental price. The red curve is the market closing price, and the blue one is the ``fundamental" price $M^j_{\;i} = n^j/n^i$. Here we took $\nu = 0.9$. } 

\section{Conclusions}
In this article we have studied a class of heterogeneous economic agent-based models which are built upon some basic set of principles, and a set of fundamental operations which include trade and product transformations. The agents have very limited information on the system, and have only a finite set of operations that they can perform.  Therefore, they have bounded rationality.  One of the key assumptions was the principle of scale invariance. That is, that there is no other scale in the economy apart from the units that agents use to measure the products, and the dynamics of the economy must be invariant under unit transformations. Moreover, all prices are relational. These assumptions are based on the idea that economic systems have approximate (or exact) gauge symmetries which restrict their possible dynamics. 

We also introduced the concept of a ``near-equilibrium" expansion. We argued that this is the analog of ``perturbation theory", which is a very familiar technique in physics. The near-equilibrium approximation allow us to derive approximate ``flow" equations that describe how many products agents exchange in a trade, and produce in a metabolic process. These equations are written for a general {\it convex} index of satisfaction. These equations can be applied to a wide variety of market models as we showed in the text.

An interesting point that we studied was the (possible) convergence towards a {\it global} economic equilibrium. In section 5.1 we saw that this can indeed happen. Moreover, we calculated the relaxation time to equilibrium in a particular model with two products. However, as we pointed out in section 5.3, one can also have steady states where the agents never quite reach a price equilibrium. 

Another interesting aspect of these models is that they can show a spontaneous transition to a monetary economy. We studied this point in section 5.4.  There, we showed that the monetary phase can be explained from a mean field theory analysis, in terms of the distribution of goods.  
Finally, in section 6 we showed that one can also address the question of stock markets and agents' expectations within our class of models.

The models presented in this article can be extended in variety of ways. First of all, one would like to develop a more systematic treatment of the near-equilibrium expansion. In particular, one would like to have a way of analytically studying the stability of certain economic equilibria and calculating relaxation times. Moreover, one would like to do this for different network topologies. It turns out, that the natural language to study such questions is that of Markov Chains and Graph Theory. This is currently work in progress \cite{Severini}. See also \cite{Wilhite} for a review on networks in economics. 

Another important question to address is the role of contracts and credit. These must be added to any realistic model of the economy. Contracts require us to go beyond anonymous agents. This is a difficult topic to deal within agent-based models. Nevertheless, it is possible to introduce contracts into our class of models. This is also work in progress.

Finally, there is the question of innovations. That is, new products come into existence quite frequently in a real economy. This question has recently be addressed in \cite{Bas}. The authors studied an economy of wealth maximizing agents. The agents were endowed with metabolic processes like the ones we have studied here. Moreover, new products can pop into existence. In their economy, though, all agents agree on a price and they could only trade using a prescribed money. Therefore, it would be interesting to extend the work of \cite{Bas} using our scale invariant approach. 

The author believes that the ultimate goal of any  non-equilibrium approach to economics, is to ultimately provide a prediction for certain {\it statistical laws} that can be matched with econometric data. These ``laws" might involve relations between the distribution of prices and the underlying products, as an example. The semi-analytical modeling approach presented here is amenable to this kind of prediction.  However, more realistic market models must be studied.

\section*{Acknowledgments}
I would like to thank Lee Smolin for introducing me to the idea that gauge invariance might play a role in economics. I would also like to thank Mike Brown, Zoe-Vonna Palmrose,  Jim Herriot and Stuart Kauffman for many interesting discussions on these topics. In fact, their ``PartEcon" model was the whole inspiration for this project. I would like to specially thank Mike Brown and Jim Herriot for inviting me to Washington, where part of this research was done. I also had many wonderful discussions with Sabine Hossenfelder, Sundance Bilson-Thompson, Philip Goyal, Laurent-Freidel, Nicolas Menicucci, Simone Severini, Yasser Omar, and Kirsten Robinson.  I also thank Kelly John and Cars H. Hommes for useful comments regarding the manuscript. Research at
Perimeter Institute is supported by the Government of Canada through
Industry Canada and by the Province of Ontario through the Ministry
of Research \& Innovation.

\end{document}